\documentclass[twocolumn]{aastex63}

\received{August 8, 2021}
\revised{\today}
\accepted{December 23, 2020}
\submitjournal{ApJ}

\shorttitle{Extreme Jellyfish Galaxies Versus Other Cluster Disk Galaxies}
\shortauthors{N. Luber et al.}

\graphicspath{{./}{figures/}}

\begin{document}

\title{GASP XXXVII: The Most Extreme Jellyfish Galaxies Compared to Other Disk Galaxies in Clusters, an HI Study}

\correspondingauthor{Nicholas Luber}
\email{nicholas.m.luber@gmail.com}

% Direct contributors.
\author{N. Luber}
\affiliation{Department of Physics and Astronomy, West Virginia University, P.O. Box 6315, Morgantown, WV 26506, USA}
\affiliation{Center for Gravitational Waves and Cosmology, West Virginia University, Chestnut Ridge Research Building, Morgantown, WV 26505}

\author{A. M\"uller}
\affiliation{Ruhr University Bochum, Faculty of Physics and Astronomy, Astronomical Institute, Universit\"atsstr.150, 44801 Bochum, Germany}

\author{J. H. van Gorkom}
\affiliation{Department of Astronomy, Columbia University, 550 West 120th Street, New York, NY 10027, USA}

\author{B. M. Poggianti}
\affiliation{INAF-Osservatorio Astronomica di Padova, Vicolo dell'Osservatorio 5, I35122 Padova, Italy}

\author{B. Vulcani}
\affiliation{INAF-Osservatorio Astronomica di Padova, Vicolo dell'Osservatorio 5, I35122 Padova, Italy}

\author{A. Franchetto}
\affiliation{INAF-Osservatorio Astronomica di Padova, Vicolo dell'Osservatorio 5, I35122 Padova, Italy}
\affiliation{Dipartimento di Fisica e Astronomia Galileo Galilei, Università di Padova, vicolo dell'Osservatorio 3, I35122, Padova, Italy}

% Contributors via comments.
\author{C. Bacchini}
\affiliation{INAF-Osservatorio Astronomica di Padova, Vicolo dell'Osservatorio 5, I35122 Padova, Italy}

\author{D. Bettoni}
\affiliation{INAF-Osservatorio Astronomica di Padova, Vicolo dell'Osservatorio 5, I35122 Padova, Italy}

\author{T. Deb}
\affiliation{Kapteyn Astronomical Institute, University of Groningen, landleven 12, 9747 AD, Groningen, The Netherlands}

\author{J. Fritz}
\affiliation{Instituto de Radioastronomía y Astrofísica, UNAM, Campus Morelia, A.P. 3-72, C.P. 58089, Mexico}

\author{M. Gullieuszik}
\affiliation{INAF-Osservatorio Astronomica di Padova, Vicolo dell'Osservatorio 5, I35122 Padova, Italy}

\author{A. Ignesti}
\affiliation{INAF-Osservatorio Astronomica di Padova, Vicolo dell'Osservatorio 5, I35122 Padova, Italy}

\author{Y. Jaffe}
\affiliation{Instituto de F\'isica y Astronom\'ia, Facultad de Ciencias, Universidad de Valpara\'iso, Avda. Gran Breta\~na 1111, Casilla 5030, Valpara\'iso, Chile}

\author{A. Moretti}
\affiliation{INAF-Osservatorio Astronomica di Padova, Vicolo dell'Osservatorio 5, I35122 Padova, Italy}

\author{R. Paladino}
\affiliation{INAF, Istituto di Radioastronomia di Bologna, Via Gobetti 101, 40129 Bologna, Italy}

\author{M. Ramatsoku}
\affiliation{Department of Physics and Electronics, Rhodes University, PO Box 94, Makhanda, 6140, South Africa}
\affiliation{INAF - Osservatorio Astronomico di Cagliari, Via della Scienza 5, I-09047 Selargius (CA), Italy}

\author{P. Serra}
\affiliation{INAF - Osservatorio Astronomico di Cagliari, Via della Scienza 5, I-09047 Selargius (CA), Italy}

\author{R. Smith}
\affiliation{Korea Astronomy and Space Science Institute (KASI), 776 Daedeokdae-ro, Yuseong-gu, Daejeon 34055, Republic of Korea}

\author{N. Tomicic}
\affiliation{INAF-Osservatorio Astronomica di Padova, Vicolo dell'Osservatorio 5, I35122 Padova, Italy}

\author{S. Tonnesen}
\affiliation{Flatiron Institute, CCA, 162 5th Avenue, New York, NY 10010, USA}

\author{M. Verheijen}
\affiliation{Kapteyn Astronomical Institute, University of Groningen, landleven 12, 9747 AD, Groningen, The Netherlands}

\author{A. Wolter}
\affiliation{INAF - Osservatorio Astronomico di Brera, via Brera, 28, 20121, Milano, Italy}

\begin{abstract}

We present the results of a VLA HI imaging survey aimed at understanding why some galaxies develop long extraplanar H$\alpha$ tails, becoming extreme jellyfish galaxies. The observations are centered on five  extreme jellyfish galaxies, optically
selected from the WINGS and OmegaWINGS surveys and confirmed to have long H$\alpha$ tails through MUSE observations. Each galaxy is located in a different cluster. In the observations there are in total 88 other spiral galaxies within the field of view (40$\arcmin$x40$\arcmin$) and observed bandwidth (6500 km s$^{-1}$). We detect 13 of these 88 spirals, plus one uncatalogued spiral, with HI masses ranging from 1 to 7 $\times$ 10${^9}$ M$_{\odot}$. Many of these detections have extended HI disks, two show direct evidence for ram pressure stripping, while others are possibly affected by tidal forces and/or ram-pressure stripping. We stack the 75 non-detected spiral galaxies and find an average HI mass of 1.9 $\times$ 10$^{8}$ M$_{\odot}$, which given their average stellar mass, implies they are very HI deficient. Comparing the extreme jellyfish galaxies to the other disk galaxies, we find that they have a larger stellar mass than almost all disk galaxies and than all HI detected galaxies, they are at smaller projected distance from the cluster center and at higher relative velocity to the cluster mean than all HI detections and most non-detections. We conclude that the high stellar mass allows extreme jellyfish galaxies to fall deeply into the cluster before being stripped and the surrounding ICM pressure gives rise to their spectacular star-forming tails.

\end{abstract}


\bigskip 

\section{Introduction} \label{sec:intro}

\quad Galaxy clusters are the largest virialized structures in the universe. The mix of galaxy populations in clusters differs from regions of lower galaxy density exhibiting the so called morphology density relation, with a much higher fraction of ellipticals and S0s in the cluster centers than elsewhere \citep{hubble31,dressler80}.  These observations have led to a nature-versus-nurture debate where galaxies are thought to be different either because they were formed in higher density regions (nature) or because the environment affects their evolution. Although this debate is still not completely settled, there is a wealth of observations that show that galaxies do get affected by their environment \citep{cortese21,boselli21}. The main processes being invoked are gas-gas interactions and gravitational interactions.

\quad The GAs Stripping Phenomena in galaxies (GASP) survey \citep{poggianti17_GASP1} is a multi-wavelength study of the processes driving the removal of gas from galaxies. In particular, GASP focuses on studying the mechanisms driving the creation of jellyfish galaxies, which have H$\alpha$ tendrils, that extend from the galaxy disks with a length at least comparable to that of the galaxy diameter. Jellyfish galaxies appear to exist in a range of environments, from the very dense centers of large clusters to the low-density outskirts \citep[e.g.][]{vulcani18b,Gullieuszik20,vulcani2021arxiv}. There is a growing consensus that many jellyfish galaxies have been shaped by ram pressure stripping, i.e. the removal of cold gas from the galaxy disk by pressure from the hot intra-cluster medium \citep{GunnGott72}. Although jellyfish galaxies are undergoing ram pressure stripping, there are also many examples of galaxies undergoing ram pressure stripping that do not show the jellyfish galaxy morphology \citep{chung09}. This brings up the question: does the jellyfish galaxy  morphology occur at a particular time in the stripping process, does it depend on the stripping parameters, or is it the local environment of the jellyfish that makes them light up in H$\alpha$? 

\quad In this paper, we explore the differences between the most extreme jellyfish galaxies, and other disk galaxies in clusters, and aim to address why some galaxies in clusters become extreme jellyfish galaxies while others do not. For this work, we define extreme jellyfish galaxies as those that the authors in \citet{poggianti16} gave a  Jclass rating of 4 or 5 to, from the homogeneous sample of observations of clusters with a redshift between 0.04 - 0.07. The visual inspection of optical images in \citet{poggianti16} found 344 candidate jellyfish galaxies in 71 clusters, 34 of which were classified as Jclass=4,5. These candidates were subsequently confirmed to have large tails of ionized gas in \citet{poggianti17_GASP1}. In order to explore the differences between these two samples, we analyze neutral hydrogen and radio continuum data for disk galaxies in clusters of galaxies that contain an extreme jellyfish galaxy.

\quad HI observations have been used extensively to study the environmental effects in clusters. Disk galaxies in clusters are HI deficient compared to galaxies in lower density environments. \citet{solanes01} show in a study of 18 nearby clusters that the HI deficiency decreases gradually with increasing projected distance from the cluster center out to about 2 Abell radii. For an indepth study of the effect of the environment on the HI scaling relations see \citet{cortese11}. To understand the physical mechanisms that affect the neutral hydrogen we can distinguish between hydrodynamical interactions, such as ram pressure stripping, turbulent viscous stripping and evaporation, and gravitational interactions, such as tidal interactions, harassment and mergers. While the former only affects the gas, the latter affects both stars and gas. Since neutral hydrogen extends to the outer parts of galaxies it is an excellent tracer of these interactions. HI imaging can help to identify which mechanism is at work. High resolution HI imaging of Virgo and Coma show many examples of this \citep{cayatte90,chung09,bravo-alfaro02,vangorkom04}. \citet{cayatte94} pioneered identifying the physical mechanisms at work by studying the detailed surface density distributions of the HI gas. In general what is found is that galaxies in the outskirts of clusters have HI disks extending beyond the optical. Then as they travel inward, the gas gets removed from one side of the disk, and often extraplanar gas and HI tails are observed. Once they reach the center, only very small symmetrical HI disks are seen, or they remain undetected \citep{chung07,chung09,jaffe15,healy20arxiv,gogate20}. The HI morphology combined with relative velocities and location in the clusters have been used to study the orbital history of the galaxies \citep{vollmer01,jaffe15,jaffe16,yoon17}.

\quad In this paper, we use both the HI morphology to identify the physical mechanisms affecting the galaxies and the projected position and relative velocity in phase space diagrams to identify the orbital history of the galaxies \citep{oman13,jaffe15,rhee17,pasquali19}. We will use the presence of an undisturbed stellar disk combined with HI removed from within the optical disk as evidence for ram pressure stripping, while a disturbed stellar disk and stellar tidal features might be a sign of a gravitational interaction. The five extreme jellyfish galaxies in our sample are J0135, J0194, J0201, J0204 and J0206 \citep{poggianti17_nat}. All are currently undergoing ram pressure stripping and the ones that have been imaged in HI (J0201, J0204 and J0206) show HI that has been removed from an undisturbed stellar disk \citep{deb2020,ramatsoku20,ramatsoku2019}. 

\quad This paper is structured in the following manner. In Section \ref{sec:obs} we detail the observations, image data properties, and source finding technique and parameters. We also describe the catalogues used for the ancillary data. In Section \ref{sec:results} we present the HI distributions of 14 HI detections and a  qualitative description of each of their HI morphologies. Additionally, we show the radio continuum morphology for 5 of the 14 galaxies, and derive the star formation rate. Lastly we show the stacked spectrum of 75 non-detected disk galaxies. Although  we do not directly detect them individually in HI, we do detect them in the stacked spectrum. In Section \ref{sec:disc} we discuss the implications of our results and in Section \ref{sec:conc} we give concluding remarks about the results of this study.

\quad In this work we assume a standard flat $\Lambda$CDM cosmology with H$_{0}$ = 70, $\Omega_{\Lambda}$ = 0.7 and $\Omega_{M}$ = 0.3.

\begin{deluxetable*}{ccccccc}[t!h!]
\tablecaption{Properties of Observed Clusters\label{tab:clusterprop}}
\tablecolumns{7}
\tablenum{1}
\tablewidth{0pt}
\tablehead
{
\colhead{Cluster Name} &
\colhead{X-Ray Center} &
\colhead{Central Redshift} &
\colhead{R$_{200}$} &
\colhead{M$_{200}$} &
\colhead{Velocity Dispersion} &
\colhead{Jellyfish\tablenotemark{a}} \\
\colhead{} & 
\colhead{J2000} &
\colhead{} &
\colhead{Mpc} & 
\colhead{10$^{14}$ M$_{\odot}$} &
\colhead{km s$^{-1}$} &
\colhead{}
}
\startdata
A3532    & 12:57:22.3 -30:21:50.4  & 0.0554 & 1.55 & 4.5 & 662.0 & JO135 \\
A4059    & 23:57:01.0 -34:45:36.0  & 0.0488 & 1.58 & 4.7 & 744.0 & JO194 \\
A85      & 00:41:50.4 -09:18:10.8  & 0.0557 & 2.02 & 9.9 & 859.0 & JO201 \\
A957     & 10:13:38.4 -00:55:33.6  & 0.0450 & 1.42 & 3.4 & 631.0 & JO204 \\
IIZW108  & 21:13:55.9 +02:33:54.0  & 0.0489 & 1.20 & 2.1 & 575.0 & JO206 \\
\enddata
\tablenotetext{a}{For more more details on the GASP jellyfish galaxies see \citet{poggianti17_nat,poggianti19a,Gullieuszik20,radovich19,vulcani18c,moretti20,vulcani20b}. For more information on JO204 see: \citet{Gullieuszik17,deb2020}. For more information on JO206 see: \citet{poggianti17_GASP1,ramatsoku2019,muller20}. For more information on JO201 see: \citet{bellhouse17,bellhouse19,george18,george19,ramatsoku20,campitiello21}}
\end{deluxetable*}

\begin{deluxetable*}{cccccc}[t!h!]
\tablecaption{Summary of Observations}
\tablecolumns{6}
\tablenum{2}
\tablewidth{0pt}
\tablehead
{
\colhead{Cluster Name} &
\colhead{Pointing} &
\colhead{Velocity Coverage\tablenotemark{a}} &
\colhead{R.M.S.\tablenotemark{b}} &
\colhead{Synth. Beam\tablenotemark{c}} &
\colhead{Galaxy Types\tablenotemark{d}} \\
\colhead{} & 
\colhead{hh:mm:ss.s dd:mm:ss.s} &
\colhead{km s$^{-1}$} &
\colhead{$\mu$Jy beam$^{-1}$} &
\colhead{$\arcsec$} &
\colhead{E:S0:S}
}
\startdata
A3532 & 12:57:04.6 -30:22:33.6 & 13593-18126 & 181 / 73 & $32.8 \times 12.8$ / $30.8 \times 9.8$ & 12:42:26 \\
A4059 & 23:57:01.0 -34:40:51.6 & 10023-15266 & 210 / 73  & $33.2 \times 13.3$ / $31.0 \times 10.8$ & 12:36:23 \\
A85 & 10:13:24.7 -00:54:21.6 & 9957-15373 & 182 / 40 & $33.0 \times 22.5$ / $14.8 \times 12.7$ & 6:18:9 \\
A957 & 00:41:30.2 -09:15:46.8 & 10575-16012 & 165 / 40 & $20.5 \times 15.9$ / $14.8 \times 13.6$& 5:14:13 \\
IIZW108 & 21:13:46.8 +02:21:18.0 & 12772-18110 & 121 / 27 & $17.0 \times 14.9$ / $14.5 \times 13.2$ & 9:22:16 \\
\enddata
\tablenotetext{a}{Velocity measurements are done using the optical barycentric definition, and the error on them is 13 km s$^{-1}$.}
\tablenotetext{b}{In a single continuum subtracted channel of channel width 26.4 km s$^{-1}$. / In the continuum image.}
\tablenotetext{c}{In the HI cube. / In the continuum image.}
\tablenotetext{d}{Breakdown of galaxy types, as defined by \citet{fasano12}, within the FWHM of the VLA primary beam and the velocity range probed for our observations.}
\label{tab:obs}
\end{deluxetable*}

\begin{deluxetable*}{cccccc}[t!h!]
\tablecaption{Summary of Data Properties of Galaxies with HI Emission}
\tablecolumns{6}
\tablenum{3}
\tablewidth{0pt}
\tablehead
{
\colhead{Detection ID\tablenotemark{a}} &
\colhead{Cluster} &
\colhead{Position\tablenotemark{b}} &
\colhead{Central Velocity\tablenotemark{c}} &
\colhead{HI Flux} &
\colhead{Continuum Flux}\\
\colhead{} & 
\colhead{} & 
\colhead{hh:mm:ss.s dd:mm:ss.s} &
\colhead{km s$^{-1}$} &
\colhead{Jy km s$^{-1}$} & 
\colhead{mJy}
}
\startdata
JO204 & A957 & 10:13:46.8 -00:54:51.1 & 12703 & 0.141$\pm$0.014 & 5.93$\pm$0.30 \\
WINGSJ101328.71-011007.6 [a] & A957 & 10:13:28.7 -01:10:07.6 & 14951	& 0.550$\pm$0.120 & $<$0.12 \\
WINGSJ101306.33-010216.6 [b] & A957	& 10:13:06.3 -01:02:16.6 & 14221 & 0.167$\pm$0.027 & $<$0.12 \\
WINGSJ101302.49-005348.7 [c]\tablenotemark{d} & A957 & 10:13:02.49 -00:53:48.7 & 13590 & 0.293$\pm$0.057 & 0.40$\pm$0.02 \\
WINGSJ101322.39-005218.3 [d] & A957	& 10:13:22.4 -00:52:18.3 & 14255 & 0.333$\pm$0.060 & 0.42$\pm$0.02 \\
WINGSJ101318.33-004957.6 [e] & A957	& 10:13:18.3 -00:49:57.6 & 14719 & 0.749$\pm$0.054 & $<$0.12 \\
WINGSJ101321.71-004652.7 [f] & A957	& 10:13:21.7 -00:46:52.7 & 14646 & 0.121$\pm$0.026 & 0.69$\pm$0.03 \\
JO206 & IIZW108 & 21:13:47.4 +02:28:34.4 & 15316 & 0.287 & 5.30$\pm$0.27 \\
WINGSJ211346.12+021420.4 [g]\tablenotemark{e} & IIZW108	& 21:13:46.1 +02:14:20.4 & 13433 & 0.119$\pm$0.039 & 0.32$\pm$0.02 \\
WINGSJ211334.23+021530.9 [h] & IIZW108	& 21:13:34.2 +02:15:30.9 & 14807 & 0.291$\pm$0.028 & $<$0.09 \\
WINGSJ211314.53+022410.7 [i] & IIZW108	& 21:13:14.5 +02:24:10.7 & 14604 & 0.121$\pm$0.024 & $<$0.09 \\
WINGSJ211404.17+022552.4 [j] & IIZW108	& 21:14:04.2 +02:25:52.4 & 16179 & 0.169$\pm$0.030 & 2.01$\pm$0.10 \\
J211330.6+022904.9 [k] & IIZW108 & 21:13:30.6 +02:29:00.3 & 15535 & 0.098$\pm$0.026 & $<$0.22 \\
JO194 & A4059 & 23:57:00.7 -34:40:50.1 & 12577 & $<$0.126 & 6.88$\pm$0.34 \\
WINGSJ235650.48-344138.2 [l] & A4059	& 23:56:50.5 -34:41:38.2	& 14568	& 0.206$\pm$0.033 &$<$0.22  \\
JO135 & A3532 & 12:57:04.3 -30:22:30.3 & 16316 & 0.121 & 2.37$\pm$0.12 \\ 
WINGSJ125623.82-301525.5 [m] & A3532	& 12:56:23.8 -30:15:25.5	& 15158	& 0.238$\pm$0.1031 & $<$0.22  \\
WINGSJ125715.40-302753.4 [n] & A3532	& 12:57:15.4 -30:27:53.4	& 16151	& 0.057$\pm$0.035 & $<$0.22 \\
JO201 & A85 & 00:41:30.3 -09:15:45.9 & 13380 & 0.113 & 4.73$\pm$0.24 \\ 
\enddata
\tablenotetext{a}{The letter corresponds to the index on Figures \ref{fig:HImorph_957}, \ref{fig:HImorph_108}, \ref{fig:HImorph_4059}, and \ref{fig:HImorph_3532}}
\tablenotetext{b}{These positions correspond to the point of peak optical emission.}
\tablenotetext{c}{Velocity measurements are done using the optical barycentric definition, and the error on them is 13 km s$^{-1}$. The central velocity is determined by taking the velocity at the middle of the HI integrated profile.}
\tablenotetext{d}{This galaxy is also identified in \citet{poggianti16} as stripping candidate JO203.}
\tablenotetext{e}{This galaxy is also identified in \citet{poggianti16} as stripping candidate JO205.}
\label{tab:results}
\end{deluxetable*}

\begin{deluxetable*}{ccccc}[t!h!]
\tablecaption{Properties of the Galaxies with HI Emission}
\tablecolumns{5}
\tablenum{4}
\tablewidth{0pt}
\tablehead
{
\colhead{Detection ID\tablenotemark{a}} &
\colhead{HI Mass} &
\colhead{Radio SFR\tablenotemark{b}} &
\colhead{Stellar Mass} &
\colhead{D$_{25}$\tablenotemark{c}}  \\
\colhead{} &
\colhead{10$^{9}$ M$_{\odot}$} &
\colhead{M$_{\odot}$ yr${-1}$} &
\colhead{10$^{9}$ M$_{\odot}$} &
\colhead{\arcsec}
}
\startdata
JO204 & $>$1.32$\pm>$0.13 & 8.82$^{+0.77}_{-0.71}$ & 75.55 & 39.0$\pm$1.7 \\
WINGSJ101328.71-011007.6 [a] & 5.15$\pm$1.12 & $<$0.18 & 0.68 & 6.2$\pm$0.4 \\
WINGSJ101306.33-010216.6 [b] & 1.56$\pm$0.25 & $<$0.18 & 3.21 & 15.4$\pm$0.4 \\
WINGSJ101302.49-005348.7 [c]\tablenotemark{d} & 2.74$\pm$0.53 & 0.59$^{+0.05}_{-0.05}$ & 7.18 & 38.0$\pm$0.8 \\
WINGSJ101322.39-005218.3 [d] & 3.12$\pm$0.56 & 0.51$^{+0.06}_{-0.05}$ & 7.45 & 15.4$\pm$0.3 \\
WINGSJ101318.33-004957.6 [e] & 7.01$\pm$0.51 & $<$0.18 & 26.17 & 21.8$\pm$0.8 \\
WINGSJ101321.71-004652.7 [f] & 1.13$\pm$0.24 & 1.03$^{+0.09}_{-0.08}$ & 4.50 & 21.8$\pm$0.3 \\
JO206 & 3.20 & 9.37$^{+0.82}_{-0.75}$ & 82.88 & 49.8$\pm$0.7 \\
WINGSJ211346.12+021420.4 [g]\tablenotemark{e} & 3.24$\pm$0.43 & 0.57$^{+0.05}_{-0.05}$ & 5.05 & - \\
WINGSJ211334.23+021530.9 [h] & 1.35$\pm$0.31 & $<$0.16 & 2.02 & - \\
WINGSJ211314.53+022410.7 [i] & 1.88$\pm$0.27 & $<$0.16 & 1.11 & 10.2$\pm$0.3 \\
WINGSJ211404.17+022552.4 [j] & 1.30$\pm$0.33 & 3.55$^{+0.31}_{-0.28}$ & 23.42 & 22.4$\pm$0.3 \\
J211330.6+022904.9 [k] & 1.09$\pm$0.29 & $<$0.16 & - & 3.4$\pm$0.1 \\
JO194 & $<$1.39 & 12.17$^{+1.06}_{-0.97}$ & 380.55 & 52.2$\pm$2.4 \\
WINGSJ235650.48-344138.2 [l] & 2.28$\pm$0.37 & $<$0.39 & 3.14 & -  \\
JO135 & $>$1.75 & 5.43$^{+0.44}_{-0.40}$ & 95.11 & 33.6$\pm$0.3 \\ 
WINGSJ125623.82-301525.5 [m] & 3.43$\pm$1.48 & $<$0.51 & 3.98 & 18.6$\pm$0.3 \\
WINGSJ125715.40-302753.4 [n] & 0.82$\pm$0.51 & $<$0.51 & 0.54 & 5.0$\pm$0.6 \\
JO201 & 1.65 & 11.01$^{+0.89}_{-0.81}$ & 92.24 & 39.2$\pm$1.1 \\ 
\enddata
\tablenotetext{a}{The letter corresponds to the index on Figures \ref{fig:HImorph_957}, \ref{fig:HImorph_108}, \ref{fig:HImorph_4059}, and \ref{fig:HImorph_3532}}
\tablenotetext{b}{Computed based on Equation 13 in \citet{SFR} assuming a spectral index between -0.7 and -0.8 with the mean value corresponding to -0.75.}
\tablenotetext{c}{This is the diameter of each galaxy at the R-band 25th magnitude isophote. With the exception of the jellyfish galaxies almost all disks are comparable in size or smaller than the 20 kpc synthesized beam of the VLA.}
\tablenotetext{d}{This galaxy is also identified in \citet{poggianti16} as stripping candidate JO203.}
\tablenotetext{e}{This galaxy is also identified in \citet{poggianti16} as stripping candidate JO205.}
\label{tab:props}
\end{deluxetable*}

\section{Data} \label{sec:obs}

\subsection{Observations and Reduction}

\quad The 5 extreme jellyfish galaxies, that were observed were taken from the sample in \citet{jellycandidates}. This sample is the first large homogeneous compilation of jellyfish galaxies in clusters from the WINGS and OmegaWINGS surveys. The five extreme jellyfish galaxies were confirmed to have large H$\alpha$ tails in follow-up MUSE observations from the GASP MUSE program. They were chosen because they are among the most convincing cases of stripping with clear tentacles of stripped ionized material. J0135 and J0194 are JClass 4 and J0201, J0204 and J0206 are JClass 5, where JClass goes from 1 to 5, with 5 showing the clearest tentacles \citep{jellycandidates}. An additional constraint was that they fell in the Declination and LST range observable with the VLA. They are located in the following clusters IIZW108, A85, A957, A3532, and A4059, which have redshifts in the range 0.045 to 0.055, see Table \ref{tab:clusterprop}. The observations were done with the Karl G. Jansky Very Large Array (VLA) in C configuration. The C-array resolution of approximately 20\arcsec, corresponding to a physical scale of approximately 20 kpc at these redshifts, is optimal for these sources, as at this resolution disk components and extraplanar gas can be distinguished, while the shortest antenna spacings are also sensitive to diffuse emission. The observations were centered spatially, and in velocity, close to, or on the location of the extreme jellyfish galaxies (See Table \ref{tab:clusterprop} for the properties of each cluster and the target extreme jellyfish galaxy). The field of view of the VLA primary beam (FWHM) is approximately 40$\arcmin$, corresponding to about 2.3 Mpc at the cluster redshifts. The velocity range probed is roughly 6500 km s$^{-1}$ (1024 velocity channels with a 31.25 kHz width). The fraction of each cluster captured in these pointings varies depending on the location and velocity of the jellyfish target and ranges from 12\% in A85, to 52\% in A957 of cluster members confirmed through spectroscopy taken in OMEGAWINGS \citep{moretti17}. Each source field was observed for a total of 20 hours, corresponding to approximately 16 hours on source, as part of JVLA project 17A-293.

\quad The data reduction was carried out using the Common Astronomy Software Application (CASA), with standard calibration methods \citep{casa}. The calibration was done in an iterative process of calculating an initial calibration, flagging Radio Frequency Interference (RFI) from the data, and recalculating the calibration on the RFI free data. The target fields were flagged manually, and with the default values of the automated flagger, \textit{RFLAG}, an internal flagging agent to CASA that flags out statistical outliers. In A3532, we had to flag approximately 15\% of the total bandwidth due to severe RFI.

\quad \textbf{Self calibration and HI imaging.} Three of the five fields had strong continuum sources which were used for several iterations of phase self calibration to reduce residual artifacts in the images. The data were smoothed to a spectral resolution of 125 kHz corresponding to a channel width of 26 km s$^{-1}$ at $z$ = 0, and the pixel size chosen was 5$\arcsec$, in order that we suitably cover the synthesized beam. We imaged the data with a robust weighting of 0.5 to create a cube with the optimal combination of resolution and sensitivity, and to have the point-spread function still be well approximated as a Gaussian. We then cleaned the data down to the 1$\sigma$ level, and identified the emission free channels. The continuum subtraction was done by considering each pixel in the image-plane independently. For each pixel, a first-order polynomial was fit to the emission free channels, and then subtracted from the entire bandwidth. This is done using the CASA data analysis task, \textit{imcontsub}. These techniques were sufficient to produce HI line cubes with little imaging artifacts and near gaussian noise. A summary of the observations and properties of the five resultant data cubes are summarized in Table \ref{tab:obs}.

\quad \textbf{Self calibration and continuum imaging.} Self calibration and imaging was repeated by transforming the already flagged and cross calibrated measurement set into Miriad \citep{miriad}. A sparse imaging was done to create a first source mask that is then used for self calibration. We iteratively shortened the solution interval and included shorter baselines into the (u,v)-range in the calibration process until a 30\,s interval was reached and all baselines were included. We re-imaged the data, improved the source masks, and repeated the self-calibration on the cross calibrated data. We created continuum images for each HI source separately in which we flagged the corresponding HI channels in the data individually to not contaminate the continuum signal with HI emission. Final images were created with a robust weighting of 0. This choice of weighting improves the resolution, in comparison to a naturally weighted image, while only having a minimal effect on the point-source sensitivity, providing us with an appropriate compromise of resolution and sensitivity (see Table \ref{tab:obs} for properties of the images). Lastly, the primary beam correction was used for all analysis and all subsequent measurements, and measurement errors, are taken from this final primary beam corrected image.

\subsection{Source Detection} \label{sec:source} 

\quad Source finding was done manually and blindly (i.e., initially ignoring any information about the optical position and redshift of galaxies in the field) through a thorough inspection of source persistence across different spatial and spectral smoothing combinations. We first required a candidate detection to have emission at, or above, the three sigma level in three adjacent channels. A candidate detection was deemed real if the emission stayed above the 2 sigma level after spatially smoothing to up to a factor 2 larger beam size, spectrally smoothing up to a factor 4, and the combination of the two. Using the deep optical images from WINGS and OMEGAWINGS \citep{fasano06,Gullieuszik15}, we only considered a detection real if an optical counterpart was seen in the images within three pixels, in the HI cube, of the peak HI emission. HI line cubes of different spatial and velocity resolutions were created and then independently searched for HI sources. Masks were created for each cube by including all emission greater than positive 2$\sigma$. These masks were combined and applied to the full resolution cubes, which were used for the final HI maps. The final HI maps were created by integrating over all pixels within the mask with values greater than positive 1.5$\sigma$. HI masses were calculated by finding the flux in a region uniform across all channels, selected by being 2 pixels wider on each side than the lowest contour on the final moment map, in order to be sensitive to any low level emission. The galaxy distances were set by the systemic velocity of the cluster (Table \ref{tab:clusterprop}).

\quad In order to understand the depth of our observations, and the galaxies we can probe, we calculate the HI observed mass limit for our data. We assume a velocity width of 132 km s$^{-1}$, three sigma emission in each channel, that the emission fills exactly one synthesized beam, and the galaxy is at the center of the pointing, and thus does not suffer from primary beam attenuation. For each cluster, we use the the distances set by the systemic velocity of the cluster (Table \ref{tab:clusterprop}) and the calculated r.m.s. in a single channel with width 26.4 km s$^{-1}$ (Table \ref{tab:obs}). This calculation yields an HI lower mass limit range  of 0.6 $-$ 1.1 $\times$ 10$^{9}$ M$_{\odot}$ for our observed clusters.

\quad We detect 14 other disk galaxies in HI emission. In Table \ref{tab:results}, we present the coordinates of the optical counterpart, HI central velocity, defined by the midpoint of the integrated profile, the HI flux, and the radio continuum flux for each galaxy. Their HI masses range from  1 to 7 $ \times 10^9 M_{\odot}$. For one of our detections no counterpart was found in the WINGS and OMEGAWINGS catalogs (see Section \ref{sec:catcompile} for a thorough description). However, inspection of the images shows a faint optical galaxy coinciding in position with the HI. 99\% of the galaxies detected by the OMEGAWINGS survey have magnitude V brighter than 24.5 \citet{Gullieuszik15}. We can therefore say that if the galaxy was not detected by the automatic source extraction, it will have a V magnitude fainter than 24.5. Since the HI line center is within the velocity range of the cluster, we assume that this optical galaxy is also a cluster member. As a result of its exclusion from the catalogues, we do not have photometry or spectroscopy for this galaxy. We therefore cannot calculate its stellar mass, and will leave it out in much of the discussion.

\quad We also searched for continuum emission from the 14 galaxies detected in HI.
Associated continuum sources were identified by overlaying the HI detections. We  identify something as a detection if the continuum flux at the position of the optical galaxy is larger than 3$\sigma$. Five of the 14 galaxies were detected in radio continuum. Upper limits are taken to be 3$\sigma$ assuming the sources are unresolved (see Table \ref{tab:props}). All five extreme jellyfish galaxies are detected in radio continuum and details about these extreme jellyfish galaxies will be presented in M\"uller et al. in prep.

\subsection{Catalogue Compilation}\label{sec:catcompile}

\quad The cluster sample is taken from  the WINGS \citep{fasano06,moretti14} and OMEGAWINGS \citep{Gullieuszik15,moretti17} surveys. A galaxy is considered a cluster member if its redshift lies within $\pm3\sigma$ from the cluster mean redshift (see \citealt{cava09}). The spectroscopic catalog is complete to a magnitude of V$\approx$ 20 mag, as defined by the Vega system. Velocity dispersions and virial radii are taken from  \citet{biviano17}. 

\quad Stellar masses were derived as in \cite{Vulcani11}, following \cite{Bell_deJong01} and exploiting the correlation between stellar mass-to-light (M/L) ratio and optical colours of the integrated stellar populations. The total luminosity, $L_{B}$, was derived from the Kron (SE{\footnotesize XTRACTOR} AUTO) observed B magnitude \citep{Varela09, Gullieuszik15}, corrected for distance modulus and foreground Galaxy extinction, and k-corrected using tabulated values from \cite{Poggianti97}. The rest-frame $(B-V)$ colour used to calculate masses was derived from observed B and V aperture magnitudes measured within a diameter of 10 kpc around each galaxy barycenter, corrected for Galaxy extinction and k-corrected as the total magnitude. Then, we used the equation $\log(M/L_{B})=a_{B} +b_{B}\times (B-V)$, for the Bruzual \& Charlot model with a \citet{salpeter55} IMF (0.1-125 M$_{\odot}$) and solar metallicity, $a_{B}$ = -0.51 and $b_{B}$ = 1.45. Then, we scale our masses to a \citet{chabrier03} IMF adding -0.24 dex to the logarithmic value of the masses. Morphological types were derived from V-band images using MORPHOT, an automatic tool for galaxy morphology, purposely devised in the framework of the WINGS project \citep{fasano12}. This catalog is complete for galaxies within the spectroscopic catalog  and only for galaxies which cover more than 200 pixels in the image. 

\begin{figure*}
\begin{center}
\includegraphics[width=\textwidth]{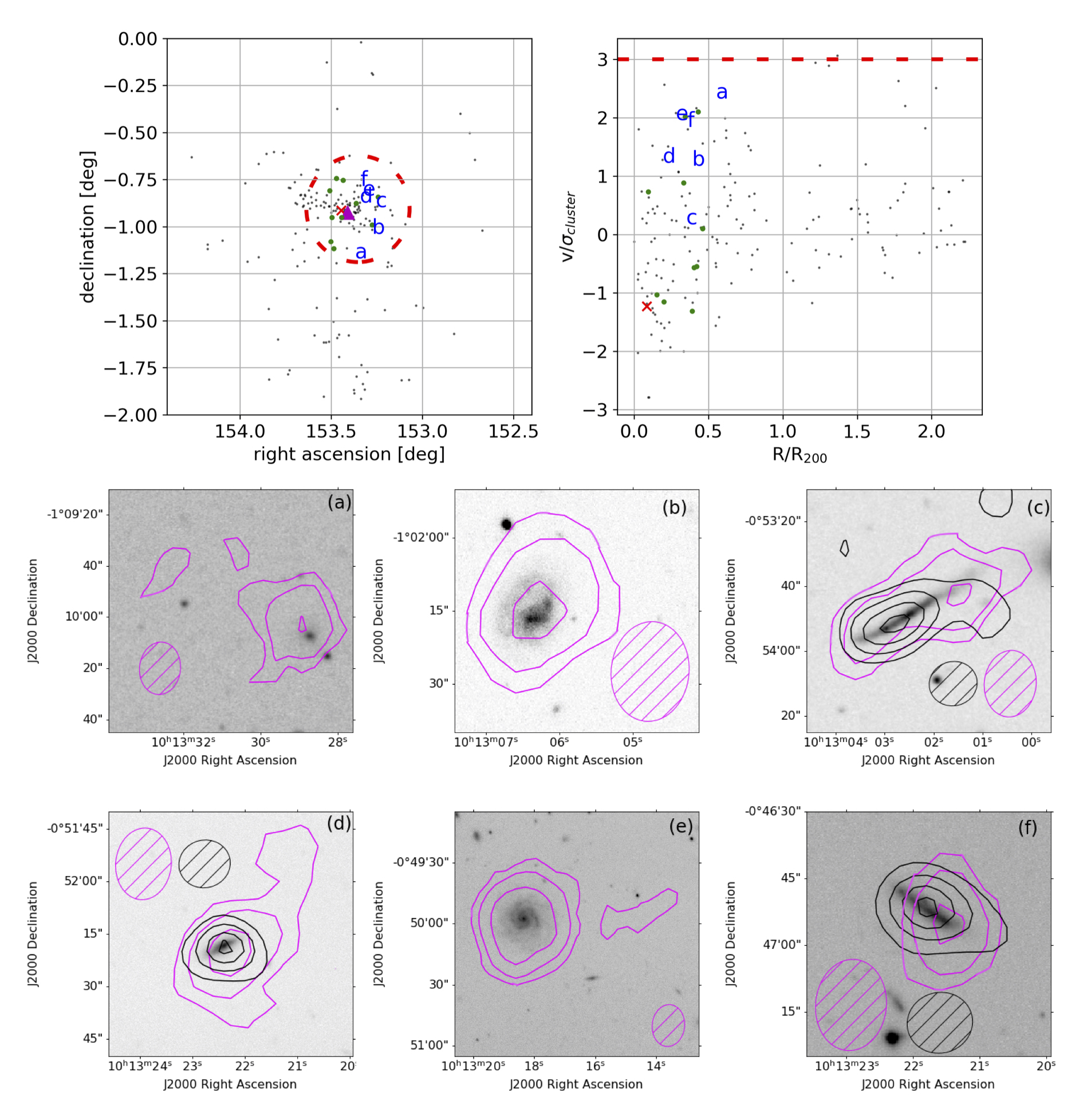}
\caption{\textbf{\textit{Top: }}The distribution of galaxies for Abell 957 in the plane of the sky (left) and phase space (right). The magenta filled triangle corresponds to the location of the brightest cluster galaxy in the cluster. The letters correspond to the locations for the correspondingly labelled detections, the red x is the location of the extreme jellyfish galaxy JO204, the black markers represent spectroscopically confirmed cluster members, and green markers represent spirals within the FWHM and velocity range of our observations. The dashed red circle in the left plot corresponds to the FWHM of the VLA primary beam for our observations. The dashed red line in the phase space diagram represents the upper limit of our observations, while the lower limit is below the boundaries of the plot. \textbf{\textit{Bottom: }}The black contours represent the radio continuum at the levels of 25, 50, 75, 90\% of the peak continuum flux for the detection, and the black hatched ellipse represents the synthesized beam for the continuum observations. The optical images are from WINGS for galaxies a and c, and from OMEGAWINGS for galaxies b, d-f. The magenta contours correspond to HI column densities of (a) $0.3 \times 10^{20} \times 2^{n}$ (b) $0.375 \times 10^{20} \times 2^{n}$ (c) $0.375 \times 10^{20} \times 2^{n}$ (d) $0.45 \times 10^{20} \times 2^{n}$ (e) $0.45 \times 10^{20} \times 2^{n}$ (f) $0.45 \times 10^{20} \times 2^{n}$  atoms cm$^{-2}$ where $n$ is equal to $1-3$. The magenta hatch ellipse corresponds to the synthesized beam for the HI data.}
\label{fig:HImorph_957}
\end{center}
\end{figure*}

\section{Results} \label{sec:results}

\quad In Table \ref{tab:clusterprop}, we list the properties of the observed clusters, their redshifts, $R_{200}$, $M_{200}$, velocity dispersions and the extreme jellyfish galaxy located in that cluster. Their masses and distances range from 2.2\,$\times$ 10$^{14}$  M$_{\odot}$ to 9.9\,$\times$ 10$^{14}$  M$_{\odot}$ and 190\,Mpc to 235\,Mpc, respectively. In Table \ref{tab:obs} we list in the last column the number of catalogued galaxies of different morphology within the FWHM of the VLA primary beam and velocity range. 

\quad In Table \ref{tab:results} we present the measured radio properties of the 14 HI detections and the five extreme jellyfish galaxies. For the 3 published extreme jellyfish galaxies the listed HI flux is taken from \citet{ramatsoku2019,ramatsoku20,deb2020}, and the continuum flux of JO206 from \citet{muller20}. One of the extreme jellyfish galaxies J0194 is not detected in HI, and we calculated an upper HI mass limit at 3$\sigma$ assuming a linewidth of 350 km s$^{-1}$, which is the stellar linewidth from \citet{poggianti17_nat}. Two other galaxies J0204 and J0135 have strong HI absorption toward the central source. The listed HI flux only includes the emission signal and thus provides a lower limit to the total HI mass. The derivation of total HI and continuum flux is described in Section 2.2. The central velocities of the HI emission are determined by taking the velocity at the middle of the HI integrated profile. In Table \ref{tab:props}, we present the HI masses, star-formation rates, derived from the continuum emission,  stellar masses and disk diameters for the 13 catalogued HI detected galaxies \citep{fasano06} and the 5 extreme jellyfish galaxies \citep{ramatsoku20,ramatsoku2019,deb2020}. 

\quad The star formation rates are calculated based on Equation 13 in \citet{SFR} using the same assumptions. We used a mean spectral index of -0.75 (with $I_{\nu}\propto\nu^\alpha$ where $I_{\nu}$ is the synchrotron surface brightness at frequency $\nu$ given the spectral index $\alpha$) for each source and estimated the errors based on a spectral index error of \textbf{$\pm 0.05$} \citep[corresponding to a spectral index range of -0.7 to -0.8][]{SFR} and the flux uncertainty given in Table \ref{tab:results}. A complication in using radio continuum to estimate star formation rates is that there may be a radioloud AGN in the center of the galaxies. Based on the BPT diagrams in \citet{poggianti19a},  Liner/AGN like emission could be identified in the central region of the extreme jellyfish galaxies. The continuum emission of the extreme jellyfish galaxies is mostly resolved and have indeed an AGN in the center. We subtracted the contribution of the radioloud AGN by making an image using only the longest antenna spacings, allowing for better resolution than the observations would normally allow, the so called super resolution. The AGN contribution was typically about $30-50$\,\%. For the five non-extreme jellyfish galaxies we were not able to identify an AGN with the emission being mostly unresolved. Our derived star formation rate for those galaxies are therefore upper limits.

\begin{figure*}[t!]
\begin{center}
\includegraphics[width=\textwidth]{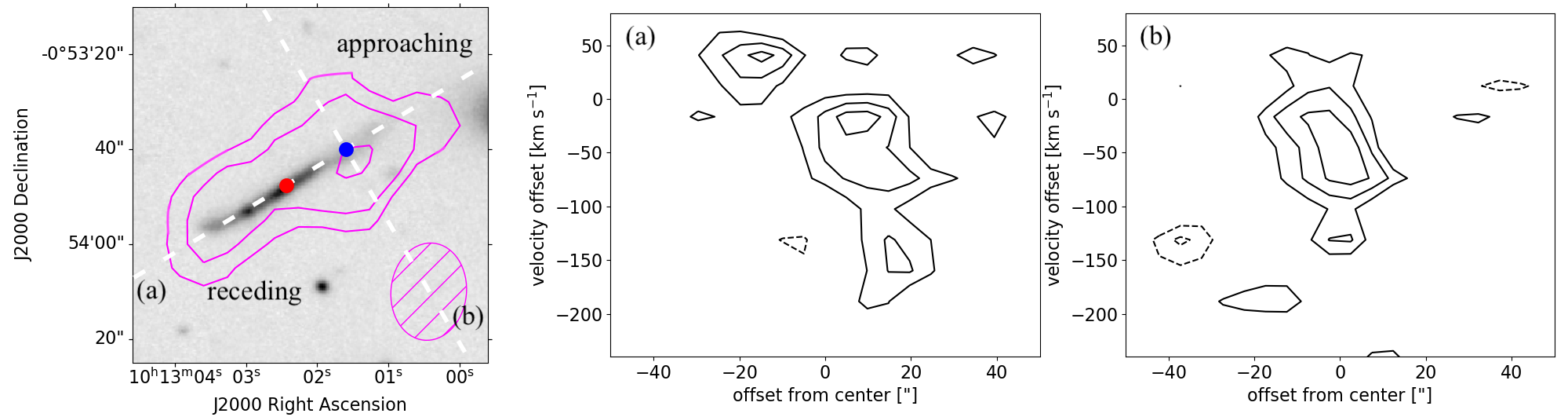}
\includegraphics[width=\textwidth]{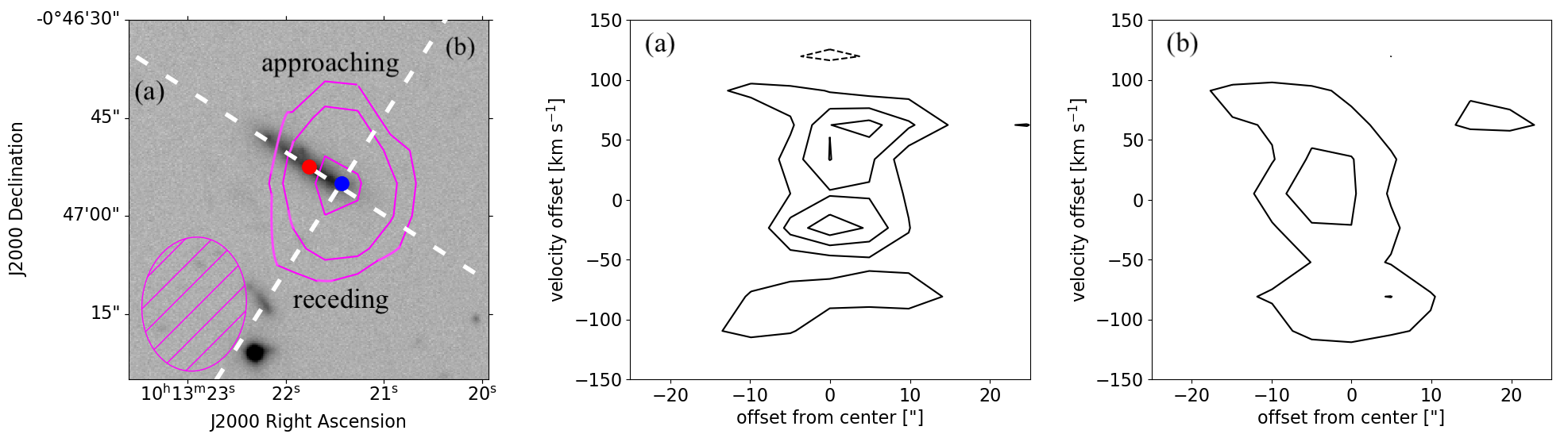}
\caption{The HI distribution and position velocity slices for galaxy c (Top) and f (Bottom). For each row, the left panel is the HI morphology overlaid on the optical morphology, with the labels approaching (the side with more negative velocities) and receding (the side with more positive velocities). The two dashed lines, labelled a and b, correspond to the axis over which the position-velocity diagrams in the middle and right panels, respectively, are taken. The slices are taken over these axes by averaging 3 pixels perpendicular to line from both above and below it, resulting in each point being the average of 7 pixels. The velocity axis is fixed to be centered on the galaxy's systemic velocity, and the spatial zero point is marked by the red marker for (a) and the blue marker for (b), with negative offset corresponding to higher right ascension. The HI morphological contours correspond to the same contours for the galaxies as in Figure \ref{fig:HImorph_957}, and the black contours correspond to $\pm$2,3,4,5$\sigma$ with dashed lines being negative, and solid lines being positive. }
\label{fig:pv957}
\end{center}
\end{figure*}

\quad In the following we present images arranged by cluster of the total HI emission and continuum emission overlaid on optical B-band images of the HI detected galaxies. HI images of J0201, J0204, J0206 can be found in \citet{deb2020,ramatsoku20,ramatsoku2019}, respectively. We present here the total HI map of J0135 and an upper limit to the HI of J0194. In these images arranged by cluster we also show projected on the sky the location of the extreme jellyfish galaxies in red, of the HI detected galaxies in blue letters, and in green the non detected catalogued disk galaxies within the FWHM of the primary beam of our VLA pointing. In addition we show for these galaxies the location in a projected distance-relative velocity phase space diagram, i.e. the projected distance from the cluster center versus the relative velocity with respect to the cluster mean velocity. In those diagrams we also indicate the velocity range probed by our observations. For a few systems we also show position-velocity diagrams along the optical and HI major axes, and selected position angles, where we considered this helpful in the interpretation.

\quad In the following subsections we qualitatively describe each individual cluster galaxy detection. A few things need to be kept in mind. While the extreme jellyfish galaxies have disk sizes (D$_{25}$ at R-band) from 34 to 50 kpc, most of the other disk galaxies have sizes of 20 kpc or less (Table \ref{tab:props}), which is comparable to the 20 kpc synthesized beam of the VLA. Most of the detections have HI extending beyond the stellar disk. We describe disturbances that we see in these outer HI disks, but we can only hint at the possible mechanisms that affect them as the signature of a gravitational or gas-gas interaction would be the same. The only direct evidence of an ICM-ISM interaction is removal of gas from the stellar disk. In some cases we lack the resolution to say this.

\subsection{A957}

\quad Abell 957 is the second smallest of the five clusters with a velocity dispersion of 631 km s$^{-1}$ and a R$_{200}$ of 1.42 Mpc. In addition to the extreme jellyfish galaxy JO204, we detect six galaxies in HI out of the 13 spiral galaxies within the primary beam. Three of the HI detections are also detected in radio continuum. Looking at the phase-space diagram in the upper right of Figure \ref{fig:HImorph_957}, we see that the non detected disks cover much of velocity space from +2$\sigma$ to -1$\sigma$. The HI detections are all at redshifted velocities with respect to the cluster mean. It could be a group falling in from the front. Of the six galaxies three, c,d, and f, show an HI morphology that could have been affected by the cluster environment. In the subsequent sections we briefly describe all detections.

\begin{figure*}
\begin{center}
\includegraphics[width=\textwidth]{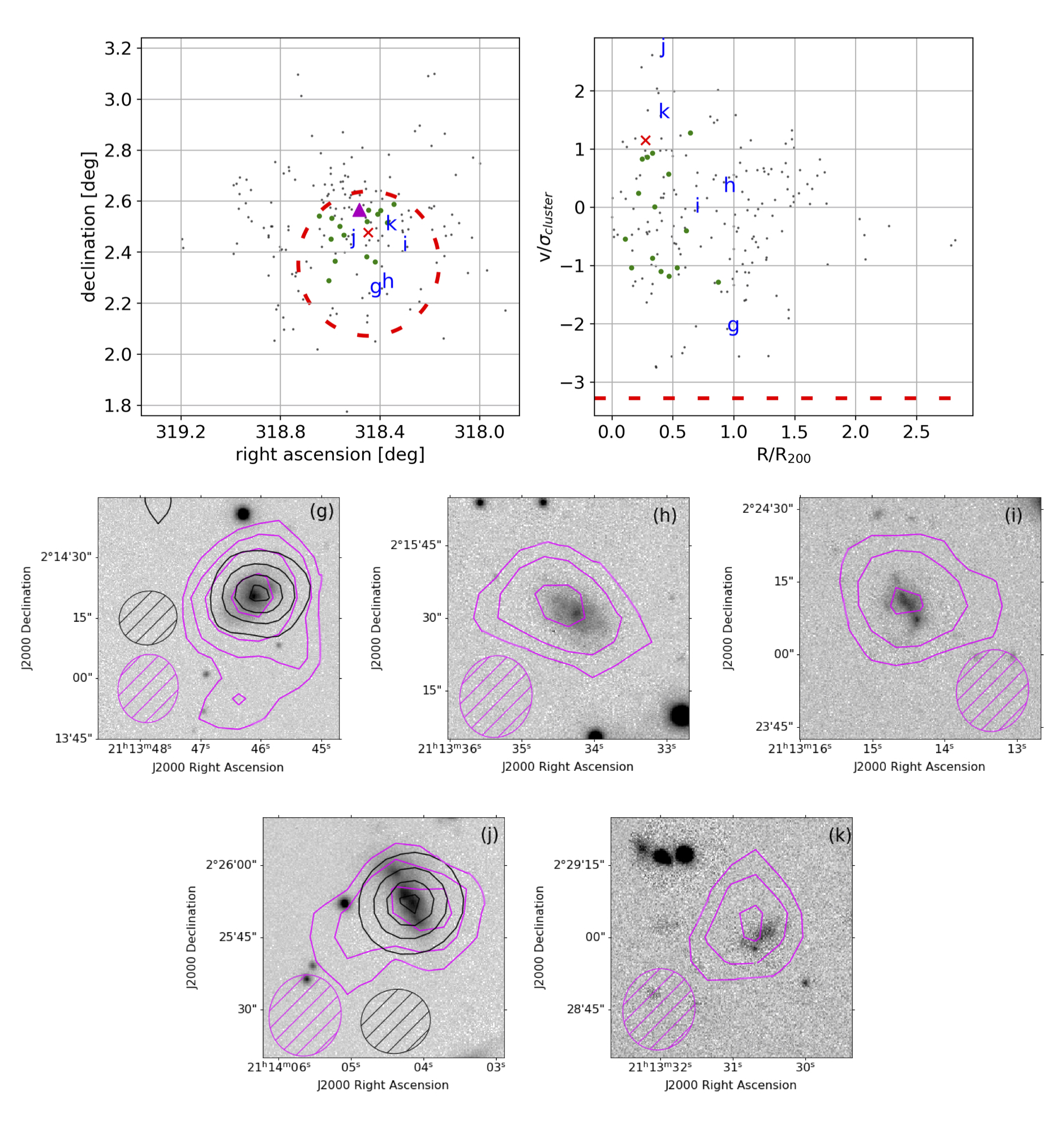}
\caption{\textbf{\textit{Top: }}The distribution of galaxies for IIZW108 in the plane of the sky (left) and phase space (right). The magenta filled triangle corresponds to the location of the brightest cluster galaxy in the cluster. The letters correspond to the locations for the correspondingly labelled detections, the red x is the location of extreme jellyfish galaxy JO206, the black markers represent spectroscopically confirmed cluster members, and green markers represent spirals within the FWHM and velocity range of our observations. The dashed red circle in the left plot corresponds to the FWHM of the VLA primary beam for our observations. The dashed red line in the phase space diagram represents the lower limit of our observations, while the upper limit is above the boundaries of the plot. \textbf{\textit{Bottom: }}The black contours represent the radio continuum at the levels of 25, 50, 75, 90\% of the peak continuum flux for the detection, and the black hatched ellipse represents the synthesized beam for the continuum observations. The optical images are from OMEGAWINGS. The magenta contours correspond to HI column densities of (g) $0.375 \times 10^{20} \times 2^{n}$ (h) $0.45 \times 10^{20} \times 2^{n}$ (i) $0.45 \times 10^{20} \times 2^{n}$ (j) $0.375 \times 10^{20} \times 2^{n}$ (k) $0.375 \times 10^{20} \times 2^{n}$  atoms cm$^{-2}$ where $n$ is equal to $1-3$. The magenta hatch ellipse corresponds to the synthesized beam for the HI data.}
\label{fig:HImorph_108}
\end{center}
\end{figure*}

\quad \textbf{WINGSJ101328.71-011007.6 (a)} The optical disk is about 6kpc, while the HI diameter is at least 40 kpc and very extended on one side (eastern direction), which is contiguous in velocity. The optical is slightly asymmetric with more diffuse light in the eastern direction as well. Some asymmetry in stars and gas is not unusual for galaxies in any environment. We do not have a redshift for the galaxy to the south west. If it turns out to be another cluster member, close in velocity to galaxy a, then it could have gravitationally disturbed the galaxy.

\quad \textbf{WINGSJ101306.33-010216.6 (b)} The HI is slightly asymmetric with respect to the optical, but is still fully enveloping the galaxy, extending at least 2 independent VLA beams, while the galaxy size of 15 kpc is less than one synthesized beam.

\quad \textbf{WINGSJ101302.49-005348.7 (c)} The optical light shows a warped disk to the east and significant debris in the northwest of the galaxy. The HI is very asymmetric extending further in the direction of the debris, the scale height increases there, and the HI peaks slightly offset from the center of the debris. This could be consistent with a warp on both sides, where the HI becomes more face on in the west. The continuum emission peaks in the center and shows very faint emission in the north-western direction. It is unclear if this galaxy's disturbed morphology is a result of ram pressure stripping or a gravitational interaction, or possibly both. The presence of stellar disturbances indicates a gravitational interaction. Furthermore, the disturbed radio continuum emission could indicate ram pressure stripping, as disturbed radio continuum emission is seen in all extreme jellyfish galaxies (see M\"uller et al. in prep.).

\quad In addition, the position-velocity diagrams (shown in the top row of Figure \ref{fig:pv957}) is suggestive of a galaxy undergoing ram pressure stripping. The HI peaks to the northwest of the optical center of the galaxy, on the approaching side of the galaxy. The large amount of total gas, and large velocity gradient in the negative velocities is likely the result of gas being removed via ram pressure stripping, and then falling to systemic velocity of the cluster. Note that the galaxy is falling in from the front and needs to decelerate in order to join the cluster.  When considering that the offset HI peak, the bend in continuum, and large gradient in velocity all occur at the same place outside of the inner region of the optical disk, the case for ram pressure stripping in this galaxy is very strong.

\quad \textbf{WINGSJ101322.39-005218.3 (d)} The galaxy shows an HI tail pointing north west, away from the cluster center and a  smaller counter tail (about half the size) to the south west. The continuum seems mostly undisturbed but has a slight elongation in the south-western direction, perpendicular to the disk. On the other hand, the stellar disk looks undisturbed. A tail and counter tail in the HI could indicate a weak gravitational disturbance by the cluster potential. There is no other galaxy nearby that could have caused this. We note that the lowest contour lies above the 2$\sigma$ expectation for the noise in the final moment map.

\quad \textbf{WINGSJ101318.33-004957.6 (e)} The HI distribution and the optical appears to be undisturbed and the HI disk is slightly more extended.  The tail to the west, although low column density is almost certainly real. As noted above, the lowest contour lies above the 2$\sigma$ expectation for the noise in the final moment map. We made HI images using different smoothing and cutoff parameters to ensure that the tail was real. The tail might have been formed in an interaction with WINGSJ101321.71-004652.7 (source f shown in Figure \ref{fig:HImorph_957}). These galaxies are very close in projection (110 kpc) and velocity (70 km s$^{-1}$). The HI in the tail is of low column density, and is pointing away from WINGSJ101321.71-004652.7.

\quad \textbf{WINGSJ101321.71-004652.7 (f)} This galaxy shows the clearest evidence for ram pressure stripping. The HI has been removed from part of the disk and peaks off center, a clear signature of ongoing ram pressure stripping. The continuum although still centered on the disk, is extended in the direction of the displaced HI, and  extends slightly beyond it. Such morphology is comparable to the findings of the extremely ram pressure stripped jellyfish galaxies \citep{muller20}.  There is however also asymmetry in the stellar disk, with more light coming from the same part of the galaxy where the HI is found. The direction of the HI points to WINGSJ101318.33-004957.6 (source e), and the two galaxies are very close in projection (110 kpc) and velocity (70 km s$^{-1}$). We can not rule out that the extended morphology is due to a gravitational interaction. 

\quad In addition, the position velocity slice (shown in the bottom row of Figure \ref{fig:pv957}) is further indicative of a galaxy undergoing ram pressure stripping. Both position-velocity slices show a large, approximately 200 km s$^{-1}$, velocity width that is being pushed from the brightest region of the optical galaxy. 

\begin{figure*}[h!t!]
\begin{center}
\includegraphics[width=\textwidth]{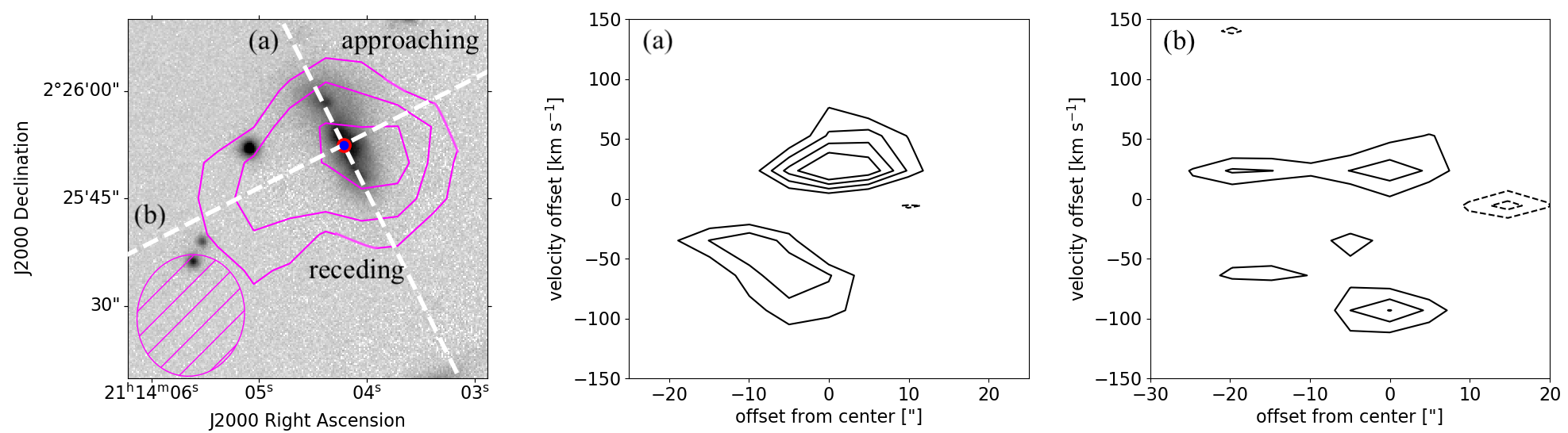}
\caption{The HI distribution for galaxy j over the optical morphology (Left), with the labels approaching (the side with more negative velocities), receding (the side with more positive velocities). The two dashed lines, labelled a and b, correspond to the axis over which the position-velocity diagrams in the middle and right panels, respectively, are taken. The slices are taken over these axes by averaging 3 pixels perpendicular to line from both above and below it, resulting in each point being the average of 7 pixels. The velocity axis is fixed to be centered on the galaxy's systemic velocity, and the position axis is centered on the red and blue marker, with negative offset corresponding to higher right ascension. The HI morphological contours correspond to the same contours for the galaxy as in Figure \ref{fig:HImorph_108}, and the black contours correspond to $\pm$2,3,4,5$\sigma$ with dashed lines being negative, and solid lines being positive. }
\label{fig:pv108}
\end{center}
\end{figure*}

\subsection{IIZW108}

\quad IIZW108 is the least massive of our five galaxy clusters with a velocity dispersion of 575 km s$^{-1}$ and a R$_{200}$ of 1.2 Mpc. In addition to the extreme jellyfish galaxy JO206 (for HI see \citet{ramatsoku2019} and continuum \citet{muller20}), we detect four HI sources out of the 16 catalogued disk galaxies. Two of the HI detections are also detected in continuum. One galaxy (k) is not catalogued. In this cluster, we see again that the catalogued disk galaxies are spread throughout the velocity range, while 4 of the 5 HI detections are at a redshifted velocity, quite possibly falling in as a group. In all cases the HI extends further than the disk. 

\quad \textbf{WINGSJ211346.12+021420.4 (g)} The HI distribution is highly asymmetric with an extension to the south of at least 20kpc beyond the optical disk, while the optical disk is asymmetric and extends further to the south east. The HI tail is pointing away from the cluster center and could be due to an interaction with the cluster potential or the onset of ram pressure stripping. Most interestingly, the continuum emission is point-like but off-center from the stellar light on the western side. 

\quad \textbf{WINGSJ211334.23+021530.9 (h)} Both the HI and the stellar distribution are symmetric. Remarkably, the HI and stellar disks are not centered on the same place.

\quad \textbf{WINGSJ211314.53+022410.7 (i)} The optical image indicates an irregular galaxy, and the HI morphology shows also an extension away from the cluster center of at least 10 kpc.

\quad \textbf{WINGSJ211404.17+022552.4 (j)} The optical morphology is complex due to a possible superposition of two galaxies. The radio continuum emission is symmetrically and point like and centered on the stellar light disk center, indicating that there is either an AGN or a bright central star-forming region. Based on another deep C-array S-band JVLA observation in full polarization mode (EVLA Project 18B-018), we find a total radio continuum intensity at S-band of 0.99$\pm$0.05 mJy, as measured from a robust 0 weighted image with an r.m.s. of 3$\mu$Jy/beam. With this, we measure a spectral index of -0.90$\pm$0.02, a polarised intensity of 0.015 mJy, and fractional polarisation of 1.5\%. This indicates that the continuum is likely comprised of both an AGN and star formation. Unfortunately, no other HI sources could be identified in the S-band data.

\quad In addition, the position velocity slice (shown in Figure \ref{fig:pv108}) is indicative of a galaxy with complex kinematics. Both position-velocity plots show a small velocity gradient along their respective axes. While the gradient along the disk is consistent with rotation, the velocities perpendicular to the disk could be due to a polar ring or disk.

\quad \textbf{J211330.6+022904.9 (k)} The HI is asymmetric about the irregular optical galaxy. Interestingly, the extension is in the direction of JO206, and the two are close in velocity ($<$20 km s$^{-1}$) and in projection ($\approx$300 kpc) indicating that this possibly due to a gravitational interaction.

\subsection{A4059}

\quad Abell 4059 is the second most massive of our five galaxy clusters   with a velocity dispersion of 744 km s$^{-1}$ and a R$_{200}$ of 1.58 Mpc. We detect only one source in HI out of 23 catalogued disk galaxies. The HI source is not detected in continuum.

\quad \textbf{WINGSJ235650.48-344138.2 (l)} The optical is clumpy but not asymmetric. The HI has an extension to the southeast, which could be caused by a gravitational interaction, although there is no close neighbour for this galaxy.

\begin{figure*}
\begin{center}
\includegraphics[scale=0.25]{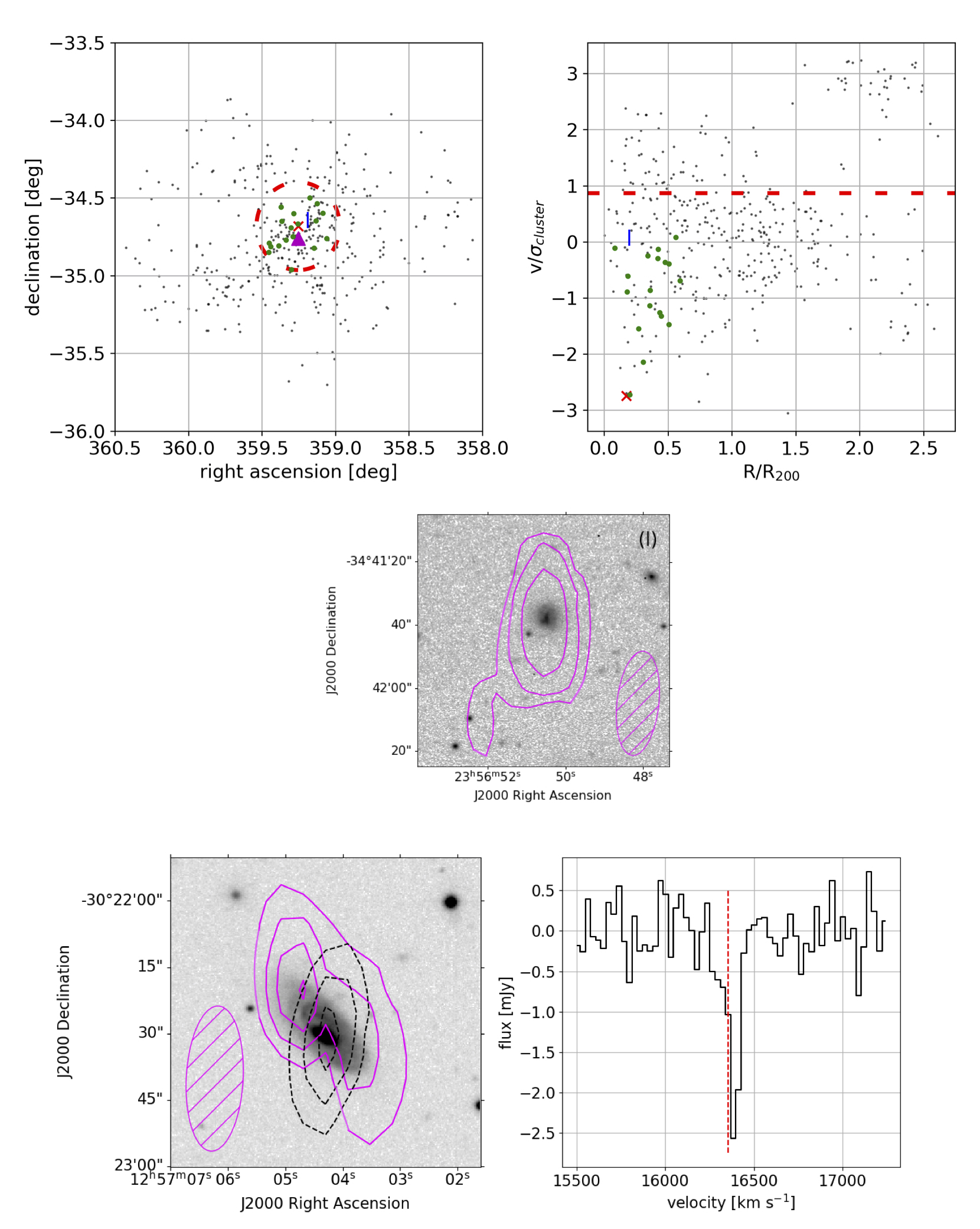}
\caption{\textbf{\textit{Top: }}The distribution of galaxies for Abell 4059 in the plane of the sky (left) and phase space (right). The magenta filled triangle corresponds to the location of the brightest cluster galaxy in the cluster. The letter corresponds to the locations for the correspondingly labelled detection, the red x is the location of extreme jellyfish galaxy JO194, the black markers represent spectroscopically confirmed cluster members, and green markers represent spirals within the FWHM and velocity range of our observations. The dashed red circle in the left plot corresponds to the FWHM of the VLA primary beam for our observations. The dashed red line in the phase space diagram represents the upper limit of our observations, while the lower limit is below the boundaries of the plot. \textbf{\textit{Middle: }}The optical image is from WINGS. The magenta contours correspond to HI column densities of $0.25 \times 10^{20} \times 2^{n}$  atoms cm$^{-2}$ where $n$ is equal to $1-3$. The magenta hatch ellipse corresponds to the synthesized beam for the HI data. \textbf{\textit{Bottom:}} \textit{Left:} The total HI absorption (black dashed contours) and HI emission (magenta contours) for the extreme jellyfish galaxy JO135. The absorption contours are at the level of -0.15,-0.1,-0.05 Jy, and the emission contours are the level of 0.5, 1.0, 1.5, 2.0 $\times$ 10$^{20}$ atoms cm$^{-2}$. The cross hatched magenta ellipse is the synthesized beam. \textit{Right:} The absorption spectral profile for JO135. The red dashed line corresponds to the systemic velocity for the galaxy.}
\label{fig:HImorph_4059}
\end{center}
\end{figure*}

\begin{figure*}
\begin{center}
\includegraphics[width=\textwidth]{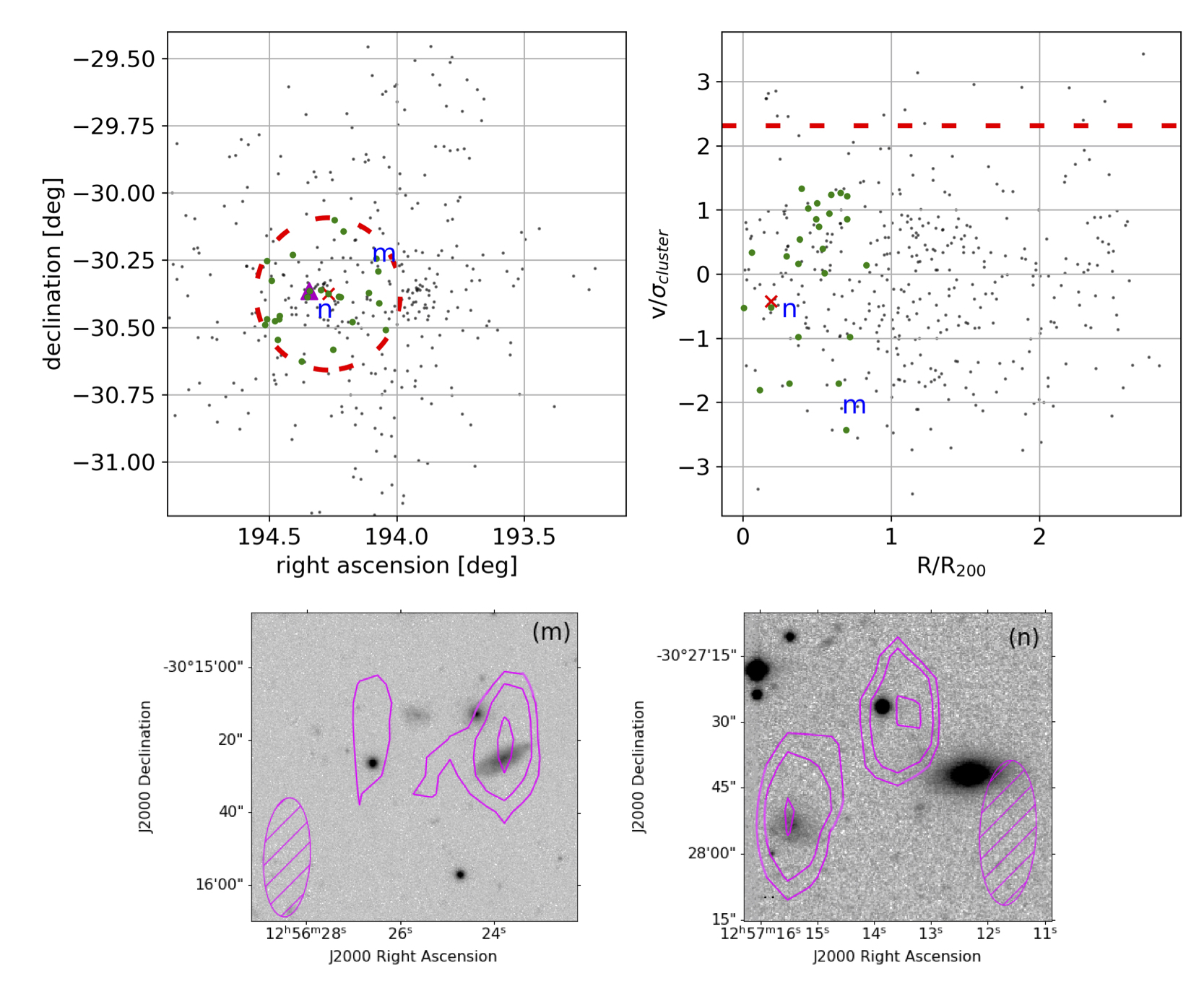}
\caption{\textbf{\textit{Top: }}The distribution of galaxies for Abell 3532 in the plane of the sky (left) and phase space (right). The magenta filled triangle corresponds to the location of the brightest cluster galaxy in the cluster. The letters correspond to the locations for the correspondingly labelled detections, the red x is the location of the extreme jellyfish galaxy JO135, the black markers represent spectroscopically confirmed cluster members, and green markers represent spirals within the FWHM and velocity range of our observations. The dashed red circle in the left plot corresponds to the FWHM of the VLA primary beam for our observations. The dashed red line in the phase space diagram represents the upper limit of our observations, while the lower limit is below the boundaries of the plot. \textbf{\textit{Bottom: }}The optical images are from OMEGAWINGS for galaxy m, and from WINGS for galaxy n. The magenta contours correspond to HI column densities of (m) $0.375 \times 10^{20} \times 2^{n}$ (n) $0.25 \times 10^{20} \times 2^{n}$ atoms cm$^{-2}$ where $n$ is equal to $1-3$. The magenta hatch ellipse corresponds to the synthesized beam for the HI data.}
\label{fig:HImorph_3532}
\end{center}
\end{figure*}

\begin{figure*}
\begin{center}
\includegraphics[width=\textwidth]{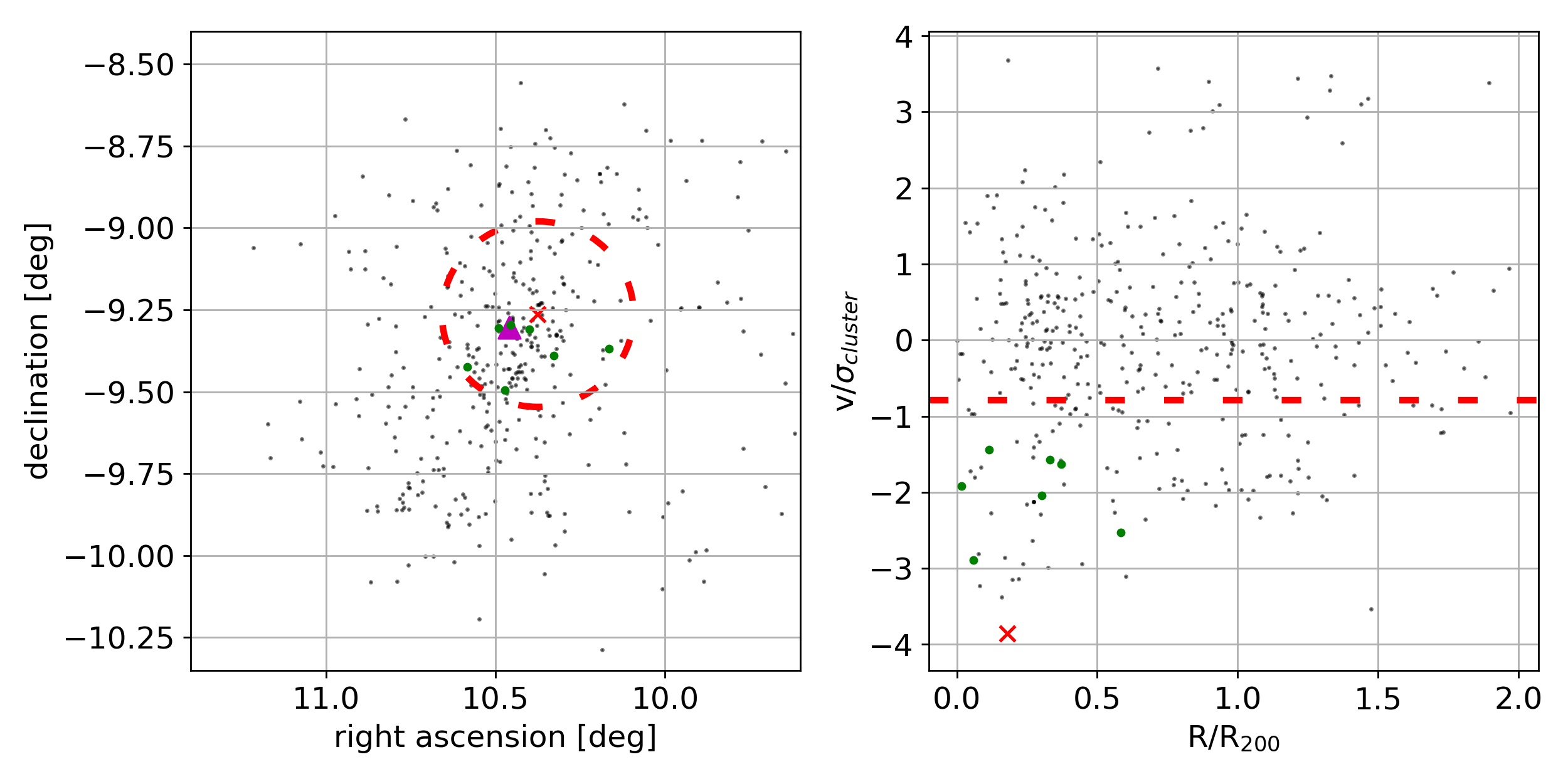}
\caption{The distribution of galaxies for Abell 85 in the plane of the sky (left) and phase space (right). The magenta filled triangle corresponds to the location of the brightest cluster galaxy in the cluster. The red x is the location of the extreme jellyfish galaxy JO201, the black markers represent spectroscopically confirmed cluster members, and green markers represent spirals within the FWHM and velocity range of our observations. The dashed red circle in the left plot corresponds to the FWHM of the VLA primary beam for our observations. The dashed red line in the phase space diagram represents the upper limit of our observations, while the lower limit is below the boundaries of the plot.}
\label{fig:HImorph_85}
\end{center}
\end{figure*}

\subsection{A3532}

\quad Abell 3532 is likely merging with the nearby cluster Abell 3530 and has a velocity dispersion of 662 km s$^{-1}$ and a R$_{200}$ of 1.55 Mpc. In addition to the extreme jellyfish galaxy JO135, we detect two additional HI sources, out of 26 catalogued disks. Neither HI detection is  detected in continuum.

\quad \textbf{WINGSJ125623.82-301525.5 (m)} The HI is asymmetric and extended in one direction. There are no known close neighbors which makes a tidal interaction unlikely for this peculiar morphology. Due to the small size of optical component of the galaxy, with respect to the VLA synthesized beam, we cannot determine if this is a case of ram-pressure stripping, because we cannot show that HI has been removed from the disk.

\quad \textbf{WINGSJ125715.40-302753.4 (n)} There is clear HI asymmetry and a cloud to the northwest. The cloud and the galaxy appear to be connected kinematically, as the velocity gradient smoothly varies. The optical galaxy is irregular, but disturbed, though not obviously enough to cause the strange HI morphology.

\subsection{A85}
\quad Abell 85 is the most massive of the five clusters. We detect no HI in A85 other than the extreme jellyfish galaxy JO201, which has been presented in \citet{ramatsoku20}.  Since the velocity of J0201 is more than 3$\sigma$ of the cluster mean we only probed the low velocity tail of the cluster velocity range, containing only 9 disk galaxies. 
 
\subsection{HI Stacking}

\quad We detect only 13 of the 88 catalogued disk galaxies within the primary beam of the observations in addition to the five extreme jellyfish galaxies. We can improve on our sensitivity by producing a stacked spectrum of the 75 non-detected spirals to calculate a mean HI mass for these 75 spirals. In order to produce this stacked signal, we smoothed each HI line cube to a common angular resolution of 35$\arcsec$ and spectral resolution of 125\,kHz. We produced a primary beam corrected flux spectrum, dividing by the dirty beam area as these sources have not been cleaned, in a region around each spiral, approximately 50\,kpc in diameter, and also calculate the primary beam corrected local noise in this region. These spectra and noise for each galaxy are then converted into units of HI mass per radio velocity channel, by using the luminosity distance to the cluster that each galaxy resides in. This allows us to average the spectra for the five different clusters, that lie at different distances, in an unbiased manner. The HI mass spectra are then averaged together channel by channel, using the local HI mass noise in each spectrum as a weight. We then subtract a first order polynomial using a fit that excludes the channels within $\pm$250 km s$^{-1}$ of line center. The HI mass is calculated by summing the values in the channels between the velocity window of -100 and +150 km s$^{-1}$. The uncertainty is calculated by using the same stacking procedure, but stacking locations with no emission, and then summing the emission in the same velocity window. This results in an average stacked HI mass for the 75 HI non detected spirals of 1.9 $\pm$ 0.4 $\times 10^{8} M_\odot$, an order of magnitude below that of our HI detections. 

\begin{figure}
\begin{center}
\includegraphics[width=\columnwidth]{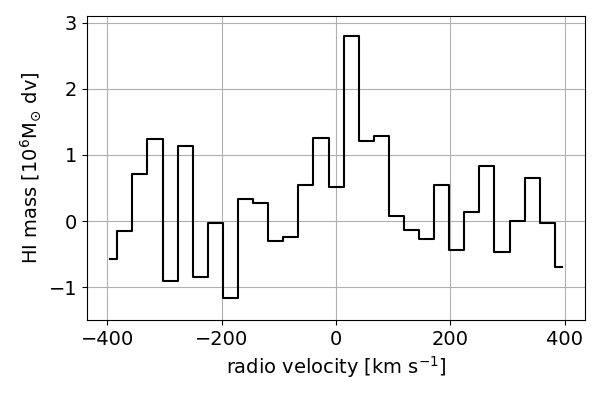}
\caption{The stacked HI spectrum for 75 spiral galaxies that were within the primary beam of our observations, but not directly detected in HI. This represents a measurement of the average HI content for the 75 spiral galaxies, and from this spectrum we measure an average HI mass of 1.9 $\pm$ 0.4 $\times 10^{8} M_\odot$.}
\label{fig:nondetect_stack}
\end{center}
\end{figure}

\quad In Figure \ref{fig:nondetect_stack}, we show the stacked spectrum of the HI emission of 75 non-detected spirals across the five clusters.

\section{Discussion} \label{sec:disc}

\subsection{HI Properties}

\quad In this section, we compare the properties of our HI detections, and non-detections of the spiral galaxies in the five clusters, with the properties of the five targeted extreme jellyfish galaxies. HI images have been published for three of the five extreme jellyfish galaxies, J0201, J0204 and J0206, and can be found in \citet{deb2020,ramatsoku20,ramatsoku2019}, respectively. In addition to their H$\alpha$ tentacles, the extreme jellyfish galaxies have HI displaced from the disk, with J0204 and J0206 showing evidence that HI is removed from within the optical disk, while J0201 still seems to have HI in the entire disk, but the authors in \citet{ramatsoku20} argue that this is a projection effect, and the HI is largely removed from the disk along the line of sight. For these three extreme jellyfish galaxies, the HI morphology confirms that these galaxies are undergoing ram pressure stripping as only a gas-gas interaction removes HI from the disk. 

\quad We show the total HI image of J0135 in the bottom row of Figure \ref{fig:HImorph_4059}. It has a radio continuum source, with integrated flux of approximately 11 mJy, in the center, against which we see HI absorption. The absorption line spectrum shown in Figure \ref{fig:HImorph_4059} is very similar to what is seen in J0204 \citep{deb2020}. The HI absorption is seen in the approaching side of the disk (the northeast of the galaxy), while the absorption is redshifted with respect to the systemic velocity, indicated by the red dotted line. This suggests that the HI is pushed toward the center via ram pressure. The HI disk is symmetrical in morphology and does not extend beyond the optical. The H$\alpha$ tail extends to the NW from the center, but unfortunately the very elongated beam in our observations makes it impossible to see if there is any HI emission from the tail, as it is in the same beam as the very strong absorption. The fifth extreme jellyfish galaxy J0194 is not detected in the current observations. 
 
\quad The HI morphologies for our 14 detections, shown in Figures \ref{fig:HImorph_957},\ref{fig:HImorph_108},\ref{fig:HImorph_4059}, and \ref{fig:HImorph_3532}, vary widely. Galaxy f in Abell 957 (\ref{fig:HImorph_957}) shows HI removed from most of the stellar disk, and is clearly undergoing ram pressure stripping. For galaxy c, the HI surface density distribution is very asymmetric, and peaks at an optically very faint extension of the disk. Although this fact by itself does not provide enough evidence for ram pressure stripping, an inspection of a position-velocity diagram (Figure \ref{fig:pv957}) at that location suggests the gas is being accelerated out of the disk toward the systemic velocity of the cluster, suggestive of ongoing ram pressure stripping. All the other detections have extended HI, with many having asymmetric extensions, but no clear hallmarks of ram pressure stripping. 

\quad In 4 galaxies, the HI asymmetry points away from the cluster center, a feature  also seen for example in the Virgo cluster \citep{chung07} and interpreted as evidence for the onset of ram pressure stripping at about the virial radius of the cluster. Unlike in Virgo where the tails extend in one direction and gas is removed from the disk on the opposite side, here we don's see direct evidence of ram pressure stripping and the tails could be due to a gravitational interaction with the galaxy cluster potential, or the onset of ram pressure stripping by the ICM. Several other galaxies are  candidates for a gravitational interaction as they have a companion that is close spatially, and in velocity. In summary, our 14 detections are being affected by the cluster environment in different ways, but due to their still high HI content, they are in the early stages of gas removal.

\quad We detect only 13 of the 88 catalogued disk galaxies, and one additionally uncatalogued galaxy. A stacked HI profile of the 75 non detected  galaxies gives an average HI mass that is an order of magnitude lower than the average mass for the 14 HI detections. In the subsequent sections, we will show that these non-detected galaxies are the galaxies that have  been most strongly affected by the environment.

\begin{figure}
\begin{center}
\includegraphics[width=\columnwidth]{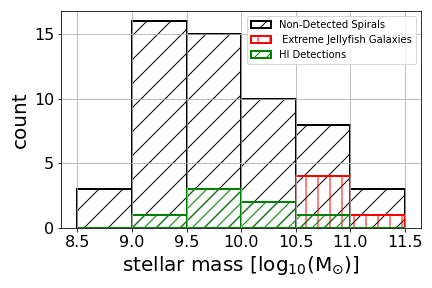}
\caption{The distribution of stellar masses for the extreme jellyfish galaxies, the HI detections, and for 75 spiral galaxies that were within the primary beam of our observations, but not directly detected. We omit one HI detection from this histogram as we were unable to secure a reliable stellar mass.}
\label{fig:stellar_mass_hist}
\end{center}
\end{figure}

\subsection{Stellar Mass of Cluster Members vs. Extreme Jellyfish Galaxies}

\quad In Figure \ref{fig:stellar_mass_hist}, we show the distribution of stellar masses for the five targeted extreme jellyfish galaxies, the 13 non-jellyfish HI detections for which we have stellar masses, and the 75 non-detected catalogued spiral galaxies. The histogram spans a stellar mass range of 10$^{8.5-12.0}$ M$_{\odot}$, with 7 bins of the equal size of 0.5dex. There is no correlation between the HI detection rate and the stellar mass of the galaxies in the cluster.

\quad The most striking result is that the 5  Jclass=4,5 extreme jellyfish galaxies have a significantly higher stellar mass than almost all spiral galaxies in the clusters. The average log stellar mass for these extreme jellyfish galaxies is 11.16 M$_{\odot}$, which is approximately a factor 5 greater than the average log stellar mass for the non-detected spirals. In addition the 5 extreme jellyfish galaxies are more massive than any of the HI detected galaxies. While \citet{jaffe18}  has found that extreme jellyfish galaxies (Jclass = 4,5)  are the most massive of all jellyfish galaxies, here we find that these jellyfish are more massive than all disk galaxies in the clusters. This brings up the possibility, does the extreme jellyfish phenomenon depend on stellar mass? 
\begin{figure*}[h!t!]
\begin{center}
\includegraphics[width=\textwidth]{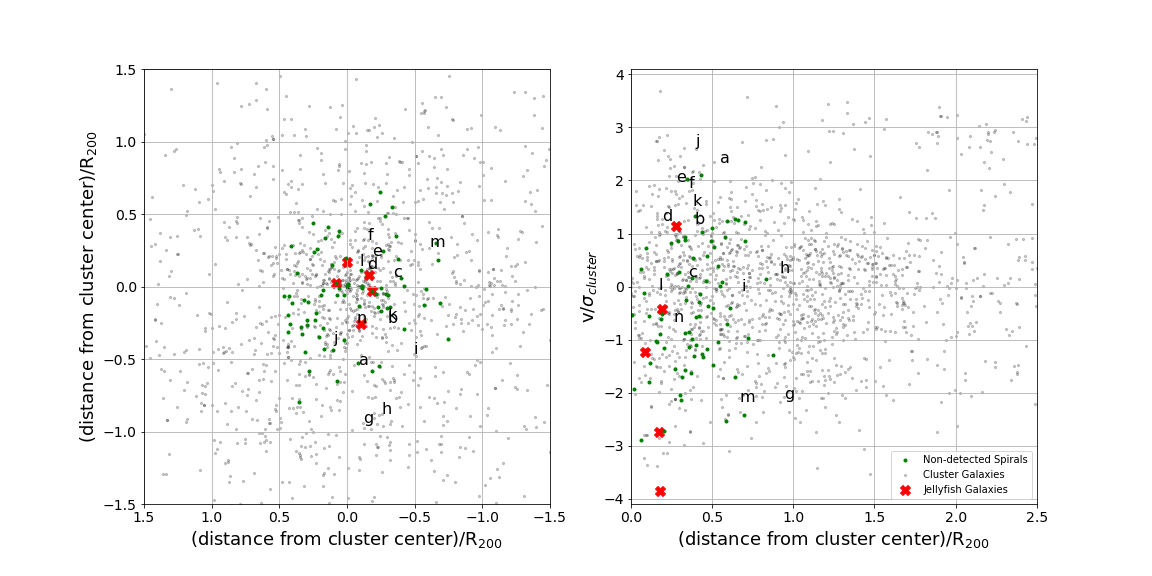}
\caption{The combined sky projection, and phase space locations for the spectroscopic members of the five observed clusters that fell within the primary beam of the VLA. The letters indicate the 14 HI detections, the red markers indicate the jellyfish, the green markers indicate the non-detected catalogued spiral galaxies, and the black markers indicate the other catalogued cluster members.}
\label{fig:combined_cluster}
\end{center}
\end{figure*}

\subsection{Location in Cluster: Extreme Jellyfish Galaxies vs. Cluster Spiral Galaxies}

\quad In Figure \ref{fig:combined_cluster}, we plot a composite of the distribution of all cluster members for the five observed clusters. On the left, we show the distribution on the plane of the sky, and on right the, projected distance - relative velocity phase space distribution. In the plot on the left, the sky coordinates have been normalized by the cluster radius (R$_{200}$) and on the plot on the right, by the cluster velocity dispersion ($\sigma$) and radius (R$_{200}$). This allows us to look at all cluster members, HI detections, spiral galaxies, and extreme jellyfish galaxies for all five clusters in a single image. 

\quad Using the information conveyed in Figure \ref{fig:combined_cluster}, we are able to distinguish the locations of the three populations we have HI information for, the stacked spirals, the directly detected spirals, and the extreme jellyfish galaxies. The mean projected radial distance, and the range of all distances in parenthesis, in Mpc for the extreme jellyfish galaxies is 0.1803 (0.0841-0.2759), for the HI detections it is 0.4671 (0.1712-0.9458), and for the stacked galaxies is 0.3915 (0.0037-0.8718). To investigate whether or not the differences in mean radial projected distances were statistically meaningful we conducted a two sample Kolmogorov–Smirnov (KS) test to compare the different combinations of our three groups. For the two combinations of the stacked spiral galaxies and the extreme jellyfish galaxies, and the HI detections and extreme jellyfish galaxies, the KS test yielded p-values of 0.00386 and 0.01152, respectively. This result confirms that the difference in mean projected distance for the extreme jellyfish galaxies, compared to the other two samples, is statistically significant. The fact that extreme jellyfish galaxies are not only in projection close, but also span a large range in relative velocities, suggests they are on radial orbits reaching their maximum velocity close to the center, as first observed in \citep{jaffe18}.

\quad What is unusual is that the extreme jellyfish galaxies still have detectable HI (in 4 out of 5 systems) despite being close (in projection) to the cluster centers, moving at high velocities, and if they are really close experiencing maximal ram pressure. \citet{jaffe18} found in an orbital study of 409 jellyfish galaxies that the most extreme jellyfish galaxies (Jclass=4,5) were at extreme locations in phase space, making it plausible that they are experiencing maximum ram pressure. Here we find that the most extreme jellyfish galaxies (Jclass=4,5) are among the most massive of all the disk galaxies, and that they are the only disks which still have significant amounts of HI at such an extreme location in phase space. Thus our results generalize the results of \citet{jaffe18}. The extreme Jclass4,5  jellyfish galaxies are more massive than most disk galaxies, and are thus more able to hold onto their neutral gas until they reach peak ram pressure in the center of the cluster. 

\begin{figure*}
\begin{center}
\includegraphics[width=\textwidth]{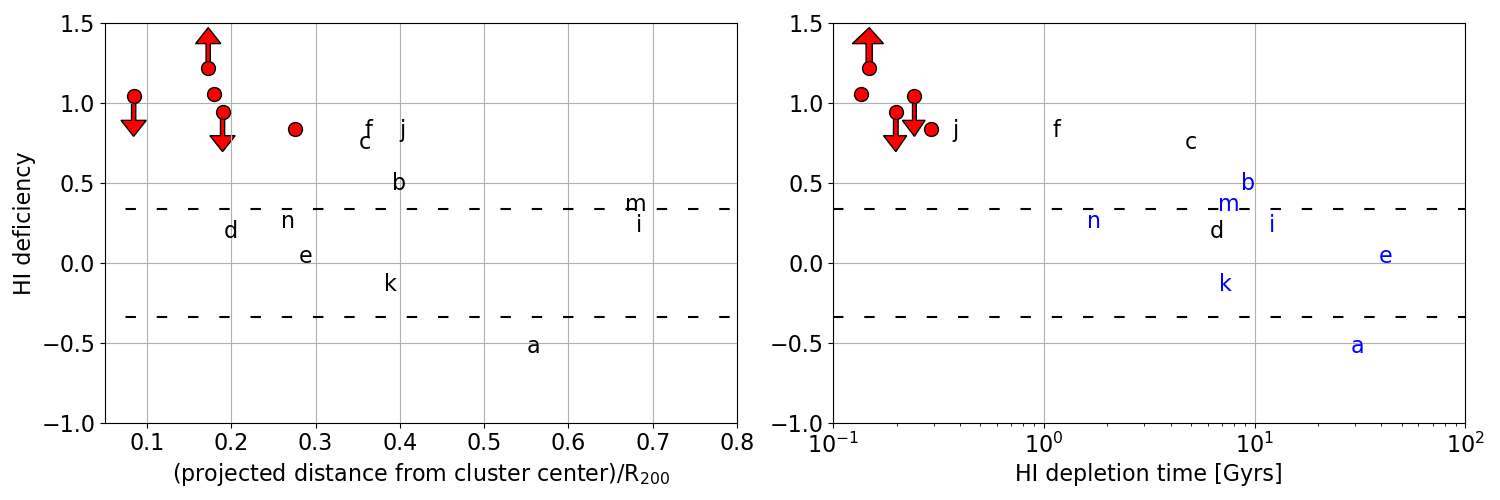}
\caption{HI deficiency as a function of \textit{Left:} the projected distance to cluster center normalized by the cluster R$_{200}$ and \textit{Right:} the HI depletion time. The red marks correspond to the jellyfish and the letters correspond to the 14 galaxies presented in this work. The two markers with down arrows indicate that the HI deficiency is an upper limit, as JO135 and JO201 have HI absorption and thus the total mass is underestimated, and the up arrow indicates the measurement is an lower limit, as we did not detect HI in JO194. Blue letters indicate values that are lower limits, based upon the upper limits of star formation measured from the radio continuum. The dashed lines  indicate the range of deficiencies considered normal for field galaxies. }
\label{fig:HIdef}
\end{center}
\end{figure*}

\begin{figure}
\begin{center}
\includegraphics[width=\columnwidth]{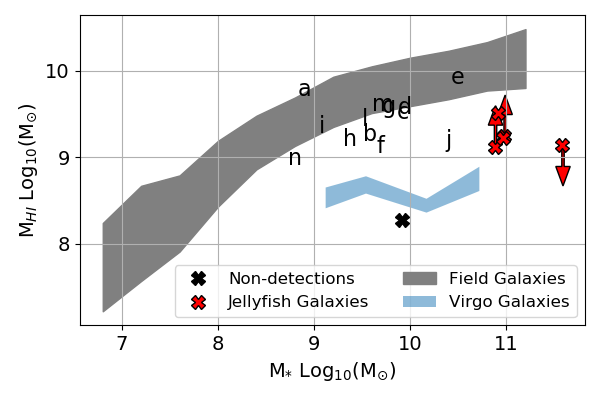}
\caption{Neutral hydrogen mass as a function of stellar mass for 5 different samples. The field galaxies are the sample from \citet{maddox15} using data from ALFALFA, and the Virgo galaxies are the sample from \citet{cortese11}. In both cases, the shaded region corresponds to the $\pm$1$\sigma$ regime which represents the scatter in HI masses for each stelllar mass bin. The stacking result is derived from the profile shown in Figure \ref{fig:nondetect_stack} for the 75 non-detected spirals. The red crosses show the values for the extreme jellyfish galaxies. The two markers with up arrows indicate that the HI mass is a lower limit, as JO135 and JO201 have HI absorption and thus the total mass cannot be calculated, and the down arrow indicates the measurement is an upper limit, as we did not detect HI in JO194. The lettered markers are the 13 detections for which we have stellar masses.}
\label{fig:stellarmass_HI}
\end{center}
\end{figure}

\subsection{HI Deficiency In Different Environments}

\quad In Figure \ref{fig:HIdef}, we show the relationship between HI deficiency and projected distance from the cluster center (left) and HI depletion time (right). HI deficiency is a useful quantitative measure of how much HI gas a galaxy should have compared to how much HI gas is measured \citet{giovanelli83}. This relationship is given by:

\begin{equation}\label{eq:hidef}
     DEF_{HI}=log_{10}M_{HI exp}-log_{10}M_{HI obs}
\end{equation}

In order to use this measure we need to first calculate the expected HI mass. This is possible via different scaling relationships that correlate a galaxy's optical  properties with HI content. For this work, we use the relationship between B-band optical diameter and HI total content as presented in \citet{denes14}. 

\begin{equation}\label{eq:exp_hi}
    log_{10}M_{HI} = 8.21 + 1.27log_{10}R_{25}
\end{equation}

We calculate the B-band R$_{25}$ for our galaxies by taking the mean semi-major axis of the elliptical isophotes, whose surface brightness values are equal to 25 mag arcsec$^{-2}$. Note that we did not correct for dust extinction. \citet{denes14} found that the scatter in the relation did not decrease using that correction, probably due to large uncertainties in correction for internal dust and inclination.  Galaxies, g,h and l are missing from the plots because no reliable diameter could be determined as they lie outside of the WINGS pointings. 

\quad Taking into account the intrinsic scatter in the HI properties of galaxies, a galaxy with an HI deficiency greater than 0.3 is HI deficient. In figure \ref{fig:HIdef}, we show that the extreme jellyfish galaxies are HI deficient, noting that the upper limits are due to the fact that there is HI absorption in those two galaxies, giving a lower limit to the total HI mass, and the HI deficiency lower limit is due to the fact JO194 had no direct detection. Three of the HI detected disk galaxies c, f, and j are HI deficient. Interestingly, of our 14 detections, f, and c, are the only galaxies with strong evidence for ongoing ram pressure stripping. Galaxy j may be a superposition of two galaxies, and if so, the optical diameter might have been overestimated. Figure \ref{fig:HIdef} recapitulates a well known fact for gas in cluster galaxies, namely  that the HI deficient galaxies  are in projection closest to the cluster center \citep{giovanelli85,solanes01,chung09,healy20arxiv}, as we also discussed in the introduction. 

\quad In order to confirm that our sample is replicating these results in a statistically significant manner, we calculate the Pearson correlation coefficient for HI deficiency and projected radius. This coefficient is then tested against the null hypothesis, that there is no correlation present. We use a t-test to calculate the probability value of the null hypothesis being true, and find that it is lower than 0.001, confirming that there is a negative correlation between HI deficiency and projected cluster radius, as was expected.

\quad  A different way to compare observed HI mass to expected HI mass is to look
at the HI mass-stellar mass relation for galaxies in different environments (\citet{cortese11}). In Figure \ref{fig:stellarmass_HI}, we compare the HI masses of the 13 detected disk galaxies, for which we also have a stellar mass, and the HI masses of the extreme jellyfish galaxies and the mean HI mass derived from the stacked spectrum of the 75  galaxies, that were not detected individually to a sample of field galaxies and a sample of galaxies in and near the Virgo cluster. The grey band shows the mean HI mass for a sample of nearby galaxies taken from \citet{maddox15}, as well as the 1$\sigma$ spread, the blue band shows the mean HI mass for a sample of galaxies in the Virgo cluster and surrounding clouds, as well as the 1$\sigma$ spread \citet{cortese11}. For both samples, the data are flux-limited. Our HI detections indicated by letters fall mostly within the grey band. In this case only galaxy f and j have somewhat less HI than expected for a galaxy in a low density environment. The mean stacked HI mass for our non detected galaxies falls on the blue line. They are as deficient as other cluster galaxies. The extreme jellyfish galaxies indicated in red are the most interesting, they are in between the two lines, on their way to losing their gas. Figure \ref{fig:stellarmass_HI} clearly demonstrates that not only are the extreme jellyfish galaxies amongst the most massive galaxies, in terms of their stellar mass, but also have a detectable amount of HI while still being deficient. This than allows the figure to serve as a synthesis of Figures \ref{fig:stellar_mass_hist} and \ref{fig:HIdef}.

\begin{figure}
\begin{center}
\includegraphics[width=\columnwidth]{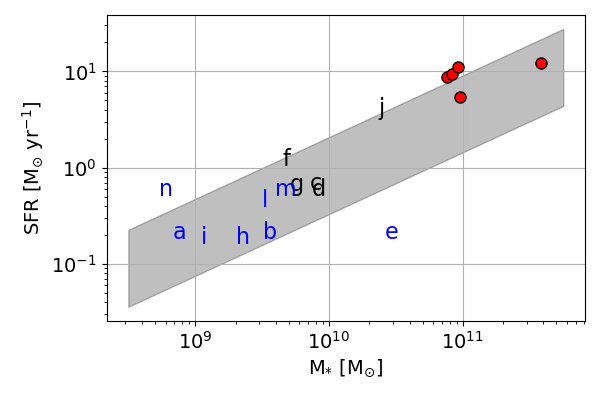}
\caption{The radio continuum star formation rate as a function of stellar mass. The letters correspond to our 13 detections, the black letters are the measured SFR rates and blue letters are the calculated upper limits. The red markers correspond to the 5 extreme jellyfish galaxies. The grey shaded region corresponds to the $\pm$1$\sigma$ regime of the SFR main sequence for cluster galaxies, as calculated in \citet{poggianti16}. The 1$\sigma$ error is measured by calculating the scatter in SFRs that independent estimators produce, which the authors in \citet{poggianti16} calculate as 0.4 dex.}
\label{fig:SFR_MS}
\end{center}
\end{figure}

\subsection{Star Formation Rates and HI Depletion Times}

\quad It has been found that jellyfish galaxies, especially those observed by GASP, have elevated star formation rates, while other spiral galaxies in clusters have star formation rates lower than those in the field \citep{vulcani18c,poggianti16}. We are able to study this relationship for our HI detections and the five extreme jellyfish galaxies (Jclass=4,5) by using the radio continuum emission to compare the star formation rates. In Figure \ref{fig:SFR_MS}, we plot the star formation rate as a function stellar mass for our 13 detections, for which we have stellar masses, and for the five extreme jellyfish galaxies. We only detect radio continuum in 5 of the 13 galaxies, and indicate this by coloring the detections black, and the non-detections blue. The shaded region is the SFR main-sequence, $\pm$1$\sigma$, as derived in \citet{poggianti16} for the WINGS sample. The SFR main-sequence for cluster galaxies is likely to be different from that for a general sample of galaxies, such as that, for example, in \citet{renzini15}. We decided to use the SFR main-sequence for cluster galaxies from \citet{poggianti16} to make sure that all quantities for our galaxies and the comparison sample were derived consistently.  In addition it offers the opportunity to see whether galaxies that are currently undergoing ram-pressure stripping differ from other cluster galaxies. Of the 5 radio continuum detected spiral galaxies (the black markers), only galaxies j and f appear to have a slightly elevated SFR for their mass compared to spiral galaxies in the WINGS sample, but still only lie at the +1$\sigma$ point. Similarly, the extreme jellyfish galaxies, J0204, J0206 and J0201 have a slightly elevated SFR. This similarity is of particular interest, as galaxies j and f are two of the three best candidates of our 14 detections that currently might undergo ram pressure stripping.

\quad With both gas and SFR in hand, we can also calculate depletion times for
the galaxies, and we show HI deficiency as a function of HI depletion time in the right panel of Figure \ref{fig:HIdef}. Interestingly, in addition to the already known fact that jellyfish have short depletion times, galaxy j and f stand out as they have both an elevated SFR and an unusually small HI reservoir. Galaxy f fits the picture of currently undergoing strong ram pressure stripping and  possibly having ram pressure enhanced star formation. We saw before that galaxy j may not be HI deficient, if in fact it is a superposition of two galaxies. The radio continuum, however, is centered on the main galaxy and the conclusion is that it is forming stars too rapidly for its HI mass. 

\quad Aside from the extreme jellyfish galaxies, we have detected 2 galaxies with stellar mass larger than 10$^{10}$ M$_{\odot}$, galaxy e and j. Do we expect these galaxies to be or become  Jclass= 4,5 jellyfish? Interestingly, galaxies e and f are at very similar location in phase space, spatially and in velocity. Galaxy f is ten times less massive than e, is HI deficient and has slightly elevated star formation. Galaxy e does show a one sided tail but there is no sign that the HI disk is experiencing ram pressure. It is not HI deficient and its SFR is unusually low. If these two galaxies are indeed physically close and experiencing similar ram pressure then it shows that stellar mass is important for a galaxy to hold onto its gas. On the other hand, Galaxy j is HI deficient, has an elevated SFR, and is already at an extreme location in phase space. It shows all the characteristics to already be, or soon become, an extreme (Jclass=4,5)  jellyfish galaxy. 

\quad Our observations are consistent with the suggestion by \citet{jaffe18} that 
galaxies with the highest stellar mass are most likely to show the extreme H$\alpha$ tentacles seen in extreme (Jclass=4,5) jellyfish galaxies. The high stellar mass enables the galaxies to hold onto their gas until they reach the highest density ICM. Although few theoretical studies have been done on star formation in tails, it has been suggested that the star formation rate depends mostly on the pressure of the ICM, for example see \citet{tonnesen12, roediger14}.

\subsection{GASP compared to other jellyfish samples, what does it mean to be extreme?}

\quad Our 5 extreme jellyfish galaxies are selected from the first large homogeneous compilation of jellyfish galaxies in clusters from the WINGS and OMEGAWINGS surveys. \citet{jaffe18} showed that within this compilation, the extreme jellyfish (Jclass=4,5) differ from the other jellyfish galaxies in the sample in stellar mass and location in phase space. We show in this paper that these extreme jellyfish galaxies differ not only from other jellyfish galaxies, but from all disk galaxies in their respective clusters in stellar mass, location in phase space, and HI properties at that location in phase space. The question is now, what does it mean to be an extreme jellyfish in GASP and how does our results compare to other surveys. A few other surveys of jellyfish galaxies have been carried out, though, we are the first to present HI properties for all disk galaxies near the jellyfish galaxies and then compare the properties of these two populations. Here we briefly discuss the differences in selection criteria and compare and contrast some results.

\quad The jellyfish galaxy candidates in the GASP survey were selected by manual inspection of each galaxy, looking for asymmetries, in the B-band images. This selection process may be biased toward actively star-forming galaxies. In fact, \citet{poggianti16} found that the star-formation rate in the jellyfish galaxies is a factor two higher than in non-stripped galaxies in this sample. Several searches for a few galaxies undergoing ram-pressure stripping, with enhanced star-formation, and long ionized tails have been performed in Virgo \citep{boselli21}, and in massive X-ray clusters at higher redshift \citep{ebeling14}. These results are consistent with ours, with \citet{ebeling14} finding that the jellyfish candidates at higher z outshine all the other disks in the clusters.

\quad \citet{yagi10,yagi17} used the Subaru telescope, and narrow band filters covering the H$\alpha$ line, to do a search for ram-pressure stripping in the nearby clusters Coma and Abell 1367. They give an in-depth discussion of the evolutionary history of the two clusters and galaxies therein. They find in both clusters spectacular tails of ionized gas, mostly, but not all, associated with galaxies. In both clusters the parent galaxies occupy extreme velocities in phase space, as we find in GASP, and the authors speculate that they are falling in for the first time.

\quad An important difference between the GASP survey and the \citet{yagi10,yagi17} surveys is that GASP is trying to find galaxies that show signs of undergoing ram-pressure stripping by looking at B-band images of the galaxies, while \citet{yagi10,yagi17} search for ionized gas tails directly with the H$\alpha$ line. \citet{yagi10,yagi17} distinguish three categories of cloud-parent connection: 1) connected H$\alpha$ clouds with disk star-formation, 2) connected H$\alpha$ clouds without disk star-formation, and 3) detached H$\alpha$ clouds. At first glance, we would expect that type 1 would be similar to the galaxies categorized in GASP as jellyfish galaxies. In Coma, all but one of the type 1 cases have parent galaxies with stellar mass in excess of 10$^{10}$ M$_{\odot}$. In Abell 1367, the stellar masses are lower than in Coma and our sample, which they suggest could be because this cluster is much less massive. This difference in cluster mass would lower the density of the ICM allowing for the extreme ram-stripping we see in jellyfish galaxies to happen to lower mass galaxies, in the manner we are hypothesize in this work. As a result, we think, in general, there is very good agreement with the surveys of \citet{yagi10,yagi17}. Their type 1 galaxies are massive galaxies currently undergoing ram pressure stripping with enhanced star formation in disk and tail. Type 2 is at a later stage of stripping, evident because the gas is already removed from the galaxy, and only some star formation is found in the tail. Type 3 galaxies have already been completely stripped, as the tails are part of the gas that has been removed from the galaxies, and, as expected, they are mostly the lower mass galaxies which get stripped first and most easily. These stripped tails without star formation would not have been detected in B-band images and  as the galaxy morphology itself is no longer affected they would not be classified as extreme jellyfish by GASP.

\section{Conclusions} \label{sec:conc}

\quad In this paper, we have presented new VLA-C array observations of neutral hydrogen in regions centered on extreme (Jclass=4,5) jellyfish galaxies in five galaxy clusters. Our work presents the HI morphology, and radio continuum morphology, for 14 non-extreme jellyfish galaxies within these clusters. In addition, we detect 75 disk galaxies that are not directly detected in HI by making a stacked spectrum. We draw comparisons between the extreme jellyfish galaxies, and the disk galaxies within the clusters. Our results are best summarized as follows:

\begin{enumerate}

    % Summary of HI results.
    \item We detect HI in 14 cluster member galaxies with a variety of HI morphologies and an average HI mass of 4$\times$10$^{9}$ M$_{\odot}$. One galaxy (f, Figure \ref{fig:HImorph_957}) shows clear evidence for ram pressure stripping, Another galaxy, c, might be affected by ram pressure (Figure \ref{fig:HImorph_957}). Most of our HI detections have HI extending beyond the stellar disk, making it difficult to  disentangle the physical mechanisms that affect them. We stacked profiles of the 75 non-detected galaxies and find an average HI mass of 1.9$\times$10$^{8}$ M$_{\odot}$.
    
    % Stellar masses.
    \item The five extreme jellyfish galaxies have a larger stellar mass than almost all disk galaxies observed in the five clusters, and they are more massive than any of our HI detected galaxies.
    
    % HI deficiency.
    \item Combining the data on stellar mass and HI content we find that our HI detections are not, or only mildly HI deficient, our non-detected spirals are as deficient as typical cluster galaxies. The Jclass=4,5 galaxies are at the extreme mass side of this diagram and have HI deficiencies in between the two populations.
    
    %Phase space
    \item The distribution in phase space is different for the extreme jellyfish galaxies than the other HI detected galaxies. Their high relative velocity and projected location close to the center makes them experience peak ram pressure.
    The less massive galaxies at these locations have already lost their gas.
    
    %Final Theory.
    \item We conclude, in agreement with \citep{jaffe18}, that the extreme jellyfish galaxies, those classified as Jclass=4,5, have such extreme star-forming tails due to their high stellar mass. The high stellar mass allows them to more easily hold onto their gas, and they only get ram-pressure stripped as they plunge through the center of the cluster. The tails of stripped gas can form stars due to the high pressure of the surrounding ICM \citep{tonnesen12, roediger14,poggianti19a}.
    
\end{enumerate}
    
\quad Our results are based on a selection of galaxies from the WINGS and OMEGAWINGS surveys and follow up MUSE data. Our selection of extreme jelly fish is based
on optical distortions of the galaxies showing signs of extraplanar light as seen in B-band images. This almost certainly means that they are currently undergoing ram pressure stripping. A next step could be to investigate whether this result applies to other ram pressure stripped galaxies with long  H$\alpha$ star forming tentacles. The stellar mass of the most extreme tail galaxies is likely to depend on the mass of the cluster, and density of the ICM. Of course there are other circumstances where the ICM pressure can locally be enhanced due to shocks resulting from infalling groups.  Future HI observations with the MeerKAT telescope of all the disks in clusters with extreme jellyfish galaxies will be useful in assessing the relation between stellar mass and location in phase space with  HI stripping, ICM pressure and star formation in stripped gas. 

\acknowledgments

We thank the anonymous referee for constructive comments that led to improvements of our presentation. 

The National Radio Astronomy Observatory is a facility of the National Science Foundation operated under a cooperative agreement by Associated Universities, Inc. J.F. acknowledges financial support from the UNAM- DGAPA-PAPIIT IN111620 grant, México. A.I., D.B., M.G., R.P., and B.V. acknowledge the Italian PRIN-Miur 2017 (PI A. Cimatti). Y.J. acknowledges financial support from CONICYT PAI (Concurso Nacional de Inserci\'on en la Academia 2017) No. 79170132 and FONDECYT Iniciaci\'on 2018 No. 11180558. MV acknowledges support by the Netherlands Foundation for Scientific Research (NWO) through VICI grant 016.130.338. Based on observations collected at the European Organization for Astronomical Research in the Southern Hemisphere under ESO programme 196.B-0578. This project has received funding from the European Research Council (ERC) under the European Union's Horizon 2020 research and innovation programme (grant agreement No. 833824). This project has received funding from the European Research Council (ERC) under the European Union’s Horizon 2020 research and innovation programme (grant agreement no. 679627). We acknowledge funding from the INAF main-stream funding programme (PI B. Vulcani).

\facilities{Karl G. Jansky Very Large Array \citep{2011ApJ...739L...1P}}
\software{Astropy \citep{2013A&A...558A..33A}, CASA \citep{casa}, MIRIAD \citep{miriad}}

\bibliography{references}{}
\bibliographystyle{aasjournal}

\end{document}